\documentclass[aip,cha,reprint,numerical]{revtex4-1}
\usepackage{amsmath}
\usepackage{amssymb}
\usepackage{graphicx}
\usepackage{natbib}		
\usepackage{color}

\def\e{\varepsilon}
\def\vp{\varphi}
\def\w{\omega}
\def\Tau{\mathcal{T}}
\newcommand{\ii}{\mathrm{i}}
\newcommand{\dd}{\mathrm{d}}

\begin{document}

\title{
A unified quantification of synchrony in globally coupled populations with the Wiener order parameter}
\author{Arkady Pikovsky}
\affiliation{Institute of Physics and Astronomy, University of Potsdam,
Karl-Liebknecht-Str. 24/25, 14476 Potsdam-Golm, Germany}
\author{Michael Rosenblum}	 
\affiliation{Institute of Physics and Astronomy, University of Potsdam,
Karl-Liebknecht-Str. 24/25, 14476 Potsdam-Golm, Germany}

\date{\today}

\begin{abstract}
We tackle the quantification of synchrony in globally coupled populations. Furthermore, we treat the problem of incomplete observations when the population mean field is unavailable, but only a small subset of units is observed. We introduce a new order parameter and demonstrate its efficiency for quantifying synchrony via monitoring general observables, regardless of whether the oscillations can be characterized in terms of the phases. Under condition of a significant irregularity in the dynamics of the coupled units, this order parameter provides a unified description of synchrony in populations of units of various complexity. The main examples include noise-induced oscillations, coupled strongly chaotic systems, and noisy periodic oscillations. Furthermore, we explore how this parameter works for the standard Kuramoto model of coupled regular phase oscillators. The most significant advantage of our approach is its ability to infer and quantify synchrony from the observation of a small percentage of the units and even from a single unit, provided the observations are sufficiently long. 
\end{abstract}

\keywords{oscillatory ensembles, synchrony, order parameter, autocorrelation, data analysis}
\maketitle

\begin{quotation}
Coupling many oscillating systems may result in a synchronization
transition, at which macroscopic collective motion appears. Examples include
pedestrians on a bridge, synchronously blinking fireflies, and applause in theater halls.
In many cases, it is not easy to distinguish synchrony and asynchrony. We suggest a method
for quantifying synchrony by measuring the regularity of the observed fields. Remarkably,
not the whole population should be observed; in many cases, just observing one unit allows for a reliable identification of the synchronization transition in the whole ensemble.
\end{quotation}

\section{Introduction}
\label{sec:intro}

The emergence of collective activity in large networks of coupled active units has been known 
for decades~\cite{Winfree-67,Kuramoto-75,Strogatz-Stewart-93,Acebron-Bonilla-DeLeo-Spigler-98,Pikovsky-Rosenblum-Kurths-01}, and yet remains a topic of intensive studies in the physics of complex systems. The relevant applications range from bridge engineering, where pedestrian synchrony under certain circumstances induces bridge vibration~\cite{Eckhardt_et_al-07},  to neuroscience, where coordination of neuronal activity forms macroscopic brain rhythms, vital or pathological~\cite{Glass-Mackey-88,Buzsaki-06,lehnertz2008epilepsy,daffertshofer2020phase}. 

This paper discusses the quantification of the degree of unit activity coordination in globally coupled ensembles, a popular model for highly interconnected networks. The addressed problem is pertinent both for theoretical and experimental studies.  A well-known solution applies when all units are accessible, so that a mean field can be obtained. Macroscopic oscillations of this mean field indicate for 
collective synchrony, while in the asynchronous case, the mean field just fluctuates. This suggest using the variance of the mean field for characterization of synchrony~\cite{ginzburg1994theory,hansel1996chaos,Pikovsky-Rosenblum-Kurths-96,golomb2001mechanisms}. A better insight is possible if the units possess well-defined phases, or the phases of all units can be inferred from data.  In that case, the  Kuramoto order parameter (KOP) and generalized Kuramoto-Daido order parameters are typically used to characterize the degree of synchrony. 

However, the quantification is not that easy in the two outlined cases.
\begin{enumerate}
\item For some systems, the phases are not defined or are difficult to extract; 
the examples detailed below include strongly chaotic systems and noise-induced oscillations, for which the phase description is hardly possible. Further relevant cases are, e.g., bursting neurons. Though the mean field observation quantifies the collective activity in these complicated cases, this measure can be substituted by a more sensitive one, as shown below.
\item The more challenging case is that of an incomplete observation. Suppose we monitor only a small subpopulation of units - can we quantify the level of synchrony in the whole, unobserved ensemble? This paper provides a positive answer for the most frequent case of a regular collective mode.
\end{enumerate}

We tackle the synchrony quantification problem by introducing a novel order parameter based on Wiener's lemma and measuring the intensity of the regular (periodic or quasiperiodic) component in a time series. 
We demonstrate that the Wiener order parameter  (WOP) (i) successfully differentiates synchronous and asynchronous states for coupled units;
(ii) in the cases where the phases are accessible and the Kuramoto-type order parameters can be calculated, the WOP still provides a better discrimination between these states than KOP and the variance-based measure for a finite-size ensemble; (iii) yields a proper quantification from an incomplete observation, even from an observation of a single unit. We illustrate our approach with examples covering noise-induced, chaotic, noisy, and regular dynamics.

The paper is organized as follows. In Section~\ref{sec:gc} we provide theoretical considerations, assumptions and limitations, define the new order parameter, and discuss the computational aspects.  
Sections~\ref{sec:ar}-\ref{sec:soq} exemplify the approach by analysis of ensembles of irregular identical and non-identical units. Section~\ref{sec:kur} extends the consideration to the case of regular  non-identical units.
Finally, Section~\ref{sec:con} concludes and discusses our findings.

\section{Regularity of mean fields and its characterization}
\label{sec:gc}

\subsection{General mean-field coupling}
\label{sec:gmfc}

We consider a population of units described by deterministic or noisy equations. We assume at the beginning that the units are identical, and the noises (terms $\vec{\xi}_k(t)$ in \eqref{eq:geq-1}), if present, are independent. Furthermore,
we assume that the systems are coupled via mean fields $\vec{Y},\vec{Z}$ which are either just global averages of some observables (global variables $\vec{Y}$) or obey dynamical equations where only the mean fields are entering (global variables $\vec{Z}$).
Then, the equations read
\begin{align}
\frac{\dd}{\dd t}\vec{x}_k&=\vec{f}(\vec{x}_k,\vec{\xi}_k(t),\vec{Y},\vec{Z}),\quad k=1,\ldots,N\;,\label{eq:geq-1}\\
\frac{\dd}{\dd t}\vec{Z}&=\vec{F}(\vec{Z},\vec{Y})\;,\label{eq:geq-2}\\
\vec{Y}(t)&=\frac{1}{N}\sum_{k=1}^N \vec{y}(\vec{x}_k(t))\;.\label{eq:geq-3}
\end{align}
Notice that dimensions of the vectors $\vec{x},\vec{\xi},\vec{Y},\vec{Z}$ generally differ. The global coupling in the population
is organized via a summation of an observable $\vec{y}$ defined as a function of local variables over all the units in the ensemble. Quite often, one assumes an ``algebraic'' coupling, where only mean fields $\vec{Y}$ are present, but in many situations, like in coupled 
clocks or metronomes on a beam~\cite{Kapitaniak_etal-12,Czolczynski_etal-13}, pedestrians on a bridge~\cite{Eckhardt_et_al-07}, and electronic oscillators with a common load~\cite{Wiesenfeld-Colet-Strogatz-98,Temirbayev_etal-12} also the global dynamical equations \eqref{eq:geq-2} are present. 

We start by assuming that individual units are irregular, i.e., their dynamics either possess noisy forces $\vec{\xi}_k$, or the units operate in a chaotic regime. Furthermore, we will assume the absence of multistability in individual units. Below, we will also discuss the potential applicability of the approach to regular systems.

In the thermodynamic limit $N\to\infty$, it is appropriate to introduce the probability density of the unit states $w(\vec{x},t)$, which evolves according to some linear operator that follows from Eq.~\eqref{eq:geq-1} and is the Liouville operator in the deterministic case, a Fokker-Plank operator in the case of Gaussian noises, or some generalized operator if the noise is not white and Gaussian. Then, instead of system (\ref{eq:geq-1}-\ref{eq:geq-3}) we can write
\begin{align}
\frac{\dd}{\dd t}w&=\hat{L}(\vec{Y},\vec{Z}) w\;,\label{eq:op-1}\\
\frac{\dd}{\dd t}\vec{Z}&=\vec{F}(\vec{Z},\vec{Y})\;,\label{eq:op-2}\\
\vec{Y}(t)&=\int \dd\vec{x}\; w(\vec{x},t)\; \vec{y}(\vec{x})\;.\label{eq:op-3}
\end{align}
Dependence of the operator $\hat{L}$ on the mean fields, which themselves depend on the density $w$ via Eqs.~(\ref{eq:op-2},\ref{eq:op-3}), makes the whole system (\ref{eq:op-1}-\ref{eq:op-3}) nonlinear. 

In this description, we attribute a steady (time-independent) solution of Eqs.~(\ref{eq:op-1}-\ref{eq:op-3}) to asynchrony and a time-dependent (periodic or more complex) solution to synchrony. Indeed, if the mean fields are constants, then the individual units can be treated as independent identical irregular oscillators. Thus, the state of each oscillator admits a statistical description with a stationary invariant distribution, which is a stable stationary solution of Eq.~(\ref{eq:op-1}).  
On the contrary, a synchronous state with oscillating mean fields can be interpreted as follows. Oscillatory mean fields impose a common forcing on all units, and due to this, the dynamics of each unit contains two components: an irregular one (and these components in different units are independent) and a forcing-induced one (and this component in different units is the same). At the calculation of the mean field, the irregular components due to the law of large numbers sum to a constant, while forcing-induced components sum to a macroscopic time-varying field.

Below, we concentrate on the most common case of regular (periodic or quasiperiodic) mean fields; a short discussion of irregular mean fields will be given in Section~\ref{sec:imf}.

\subsection{Complete and partial observations}
\label{sec:cpo}

We will suppose that a scalar observable of a specific system $u(\vec{x}_k)$ is available, possibly not for all units (such a situation often occurs in neuroimaging~\cite{wenzel2021identification}). 
Let us define the partial mean fields as
\begin{equation}
U_M(t)=M^{-1}\sum_{k\in\mathcal{S}_M} u(\vec{x}_k(t)) \;.
\label{eq:pmf}
\end{equation}
Here, $1\leq M\leq N$ is the size of the set $\mathcal{S}_M$ over which the averaging is performed. The set of the ``observed'' units $\mathcal{S}_M$ can be chosen at random; because we assume identical elements, the statistical properties of $U_M$ should not depend on this choice. The case $M=1$ means the observation of just one oscillator from the population; the case $M=N$ yields the usual mean field to which all the units contribute. We denote this full observable $U(t)$, dropping the index. Its definition is similar to that of mean fields $\vec{Y}$ (Eq.~\eqref{eq:geq-3}), and in fact one can (but this is not necessary) choose $U$ as one of the components of $\vec{Y}$. 

In the thermodynamic limit,
\begin{equation}
U(t)=\int \dd\vec{x}\; w(\vec{x},t)\; u(\vec{x})\;.
\label{eq:pmf1}
\end{equation}
In this limit, $U=\text{const}$ in an asynchronous state; while time-dependent $U(t)$ (typically periodic or quasiperiodic) 
manifests synchrony in the population. A straightforward
way to distinguish these cases is to calculate the variance of $U(t)$:
\[
\text{Var}(U)=\langle (U(t)-\bar{U})^2\rangle,\qquad \bar{U}=\langle U(t)\rangle\;,
\]
where the brackets denote statistical averaging~\footnote{In this paper we assume ergodicity and stationarity, so practically the averaging is performed over time.}. 
This quantity vanishes in the asynchronous regime and is finite in the synchronous one, thus it is widely used as an empirical order parameter characterizing transition to synchrony \cite{ginzburg1994theory,hansel1996chaos,Pikovsky-Rosenblum-Kurths-96,golomb2001mechanisms}.

\subsection{Finite-sample fluctuations and their elimination}
\label{sec:elim}

Even in the thermodynamic limit,  for partial observations of sample size $M$, one cannot expect complete regularity
 of the observations $U_M$. 
 Indeed, a scalar observable of a specific unit $u(\vec{x}_k(t))$ contains a regular part $u_{reg}(t)$ due
 to a regular forcing of this unit by the regular mean fields $\vec{Y}(t),\vec{Z}(t)$, and an irregular part $u_{irr,k}(t)$ from the irregular specific dynamics. Noteworthy is that the regular part is the same for all oscillators (here, the assumption of global coupling is important!), while irregular parts can be considered independent. In the asynchronous case, $u_{reg}(t)=\text{const}$. Now, the finite-sample average can be represented as
 \begin{equation}
U_M(t)=\frac{1}{M}\sum_{k\in\mathcal{S}_M} u_{irr,k}(t)+u_{reg}(t)\;.
\label{eq:pmf2}
\end{equation}

 In finite-sample observations $U_M(t)$, finite-size fluctuations due to the first term on the r.h.s. of Eq.~\eqref{eq:pmf2}
 may be significant
for small samples. These fluctuations lead to a non-vanishing variance $\text{Var}(U_M)$ in the whole range of parameters, also for parameters where asynchrony occurs. Here, one generally expects  a law of large numbers $\text{Var}(U_M)\sim M^{-1}$ to be valid (cf. Refs.~\onlinecite{ginzburg1994theory,hansel1996chaos,golomb2001mechanisms}), although there might be more complex dependencies in a vicinity of the synchronization transition.

We suggest eliminating the fluctuating term in Eq.~\eqref{eq:pmf2} by virtue of the autocorrelation analysis. 
We will rely on the irregularity of the local dynamics, which ensures that the autocorrelations of $u_{irr,k}(t)$ 
decay to zero at large time lags. 
For the observables $U_M(t)$ we define the standard autocorrelation functions (ACF), 
\begin{equation}
\Gamma [U_M](\tau)=\langle(U_M(t)-\bar{U}_M)(U_M(t+\tau)-\bar{U}_M)\rangle\;,
\label{eq:acf0}
\end{equation}
where $\bar{U}_M$ is the time average of the observable $U_M$. Note that $\Gamma [U_M](0)=\text{Var}(U_M)$.
Because the irregular and regular components of $U_M(t)$ can be considered as independent, the ACF \eqref{eq:acf0}
decomposes in two parts~\footnote{In the calculation of the ACF it is usual to assume that the regular component possesses a 
random phase shift}:
\begin{equation}
\Gamma [U_M](\tau)=\frac{1}{M}\Gamma_{irr}(\tau)+\Gamma_{reg}(\tau)\;,
\label{eq:acf2}
\end{equation}
where 
\[
\Gamma_{irr}(\tau)=\langle (u_{irr}(t)-\bar{u}_{irr})(u_{irr}(t+\tau)-\bar{u}_{irr}), 
\]
\[
\Gamma_{reg}(\tau)=\langle (u_{reg}(t)-\bar{u}_{reg})(u_{reg}(t+\tau)-\bar{u}_{reg})\;.
\]
Note that in the asynchronous case $\Gamma_{reg}(\tau)=0$. 

Now we utilize different time lag dependence of two parts of $\Gamma[U_M](\tau)$: while the value of $\Gamma_{irr}(\tau)$
decays at $\tau\to\infty$, the regular part $\Gamma_{reg}(t)$ is a periodic or quasiperiodic function of $\tau$.
In the power spectrum, the regular component corresponds to delta-peaks, while the irregular component corresponds to a continuous spectral density.  A proper way to quantify the intensity of the regular component in the autocorrelation function  
$\Gamma(\tau)$  is delivered by the Wiener formula~\cite{wiener1930generalized}
\begin{equation}
W(\Gamma)=\lim_{\Tau\to\infty}\frac{1}{\Tau}\int_0^\Tau \Gamma^2(\tau)\,\dd\tau\;.
\label{eq:wf}
\end{equation}
We will call $W$ the Wiener order parameter (WOP). One can easily see that because the integral $\int_0^\infty \Gamma_{irr}^2(\tau)\dd\tau$ converges, only the regular component contributes to $W$. If this regular component vanishes, as it happens in an asynchronous regime, then the WOP vanishes as well.

In the modern literature, expression \eqref{eq:wf} is often referred to as Wiener's lemma~\cite{kornfeld1982ergodic,queffelec2010substitution}. The following simple calculation supports its validity.
We are interested in the discrete component of the power spectrum, which can be represented as a sum of delta-functions
\[
S_d(\Omega)=\sum_k A_k\delta(\Omega-\Omega_k),\qquad \Gamma_d(\tau)=\sum_k A_k\cos(\Omega_k\tau)\;.
\]
Integrating the square of the ACF, we get the Wiener formula
\[
\lim_{\Tau\to\infty}\frac{1}{\Tau}\int_0^\Tau \Gamma_d^2(\tau)\,\dd\tau=\frac{1}{2}\sum_k A^2_k\;.
\]

Some remarks about practical evaluation of the WOP are in order. First, typically the ACF \eqref{eq:acf0} is calculated from a long time series $0\leq t\leq T$  as an integral
\begin{equation}
\Gamma [U_M](\tau)=\frac{1}{T-\tau}\int_0^{T-\tau} (U_M(t)-\bar{U}_M)(U_M(t+\tau)-\bar{U}_M)\,\dd t\;.
\label{eq:acf}
\end{equation}
Thus, an additional parameter, the averaging time $T$ becomes relevant.
Second, formally Eq.~\eqref{eq:wf} requires an integration over an infinite interval of time lags, so that the contribution of  the irregular component  $\Gamma_{irr}(\tau)$ vanishes due to the factor $1/\Tau$. Practically, we calculate the WOP by starting integration not at zero, but at a time lag $\Tau_1$:
\begin{equation}
W(\Gamma)=\frac{1}{\Tau_2-\Tau_1}\int_{\Tau_1}^{\Tau_2}  \Gamma^2(\tau)\,\dd\tau\;.
\label{eq:wf1}
\end{equation}
One should take $\Tau_1$ large enough so that in the interval $\Tau_1<\tau<\Tau_2$, the irregular component $\Gamma_{irr}$ is very small. 
Additionally, one has to take the interval $\Tau_2-\Tau_1$ much larger than the characteristic period of the ACF (the corresponding error can be reduced by using an appropriate windowing).
We expect finite-observation errors for finite values of $T,\Tau_2-\Tau_1$, hampering a perfect disentanglement of regular and irregular components.

Summarizing, by virtue of the ACF calculations, we eliminate (not completely, but significantly) the irregular component in the observed
field $U_M(t)$. Thus, we expect the Wiener order parameter $W$ value calculated according to the suggested procedure to be small in the irregular state. Here, only the finite-sample-induced irregular component is present in $U_M(t)$, which leads to an ACF $\Gamma_M(\tau)$ that tends to zero at large time lags $\tau$ (practically, one observes a decay to some level which depends on the averaging time $T$ used for calculation of the ACF). Effectively, via the ACF calculation, we perform an additional averaging of $U_M(t)$ in time, which can be considered an equivalent of enlarging the sample size $M$. 

In fact, dependence of the remnant fluctuations of the ACF on the averaging time $T$ in \eqref{eq:acf} can be used for a better
distinguishing of synchrony and asynchrony (cf.~Refs~\onlinecite{ginzburg1994theory,hansel1996chaos,golomb2001mechanisms}, which discuss
in a similar spirit exploration of the $M$-dependence). One can calculate the ACF for different averaging times $T$. Because the remnant fluctuations are expected to be $\sim T^{-1/2}$, the corresponding WOP value $W$ will scale as $W\sim T^{-1}$. If one observes such a scaling, then ``by extrapolation,'' one can attribute $W=0$ and conclude that the observed regime is asynchronous. In contradistinction, if the values of $W$ saturate starting from some $T$, this indicates a finite value of $W$. One thus obtains a variant of a $0-1$ test for collective synchrony.

We stress here that the case $M=1$  can also be included in the consideration. Here, one does not have any sampling-induced irregularity, and the improvement via the ACF analysis will be significant only if there is an internal irregularity of the individual oscillators. We will show below in Sections~\ref{sec:ar},\ref{sec:ca},\ref{sec:soq} the examples of noisy and chaotic individual oscillators, where even observations from one unit will provide a reasonable characterization of the synchronization transition. However, observations of a few units deliver a rather poor performance for regular oscillators in the deterministic Kuramoto model (Section~\ref{sec:kur}).

\subsection{Finite-size effects}
\label{sec:fse}

Above, we assumed that the mean fields are strictly regular, as in the thermodynamic limit; see Eqs.~(\ref{eq:op-1}-\ref{eq:op-3}).
If, instead, a finite ensemble is explored, then the finite-size fluctuations are inevitable~\cite{Daido-90,Pikovsky-Ruffo-99,bertini2014synchronization,Hong-15,gottwald2017finite,Peter-Pikovsky-18,yue2023stochastic}, especially in the asynchronous state and close to the synchronization threshold (a strongly synchronous regime can be perfectly regular even for finite ensembles). 
If the fields $\vec{Y},\vec{Z}$ in Eqs.~(\ref{eq:geq-1}-\ref{eq:geq-3}) weakly fluctuate, the same can be expected for the
complete observable $U(t)$. These basic fluctuations will be complemented with finite-sample fluctuations for all partial observables $U_M(t)$.
For us most important is that the regularity of  $\vec{Y},\vec{Z},U$ is no more perfect: while these fields are perfectly periodic in the thermodynamic limit, for finite $N$ one expects some diffusion of the collective phase (diffusion constant $\sim N^{-1}$), which means existence of a finite 
coherence time of the mean fields $t_{cor}\sim N$. Thus, the peaks in the spectrum of the mean field fluctuations $U(t)$ are no more delta-peaks but have finite width $\sim 1/t_{cor}$, and the ACF of $U(t)$ will decay at time $t_{cor}$. 

Nevertheless, for large $N$, one still can separate the spectrum into a broad-band component and narrow peaks; in terms of the ACF $\Gamma(\tau)$, one should consider an interval of time lags $\Tau_1,\Tau_2$ such that $\Tau_1$ is larger than the typical correlation time of the broad-band component, and $\Tau_2$ is smaller than $t_{cor}$. Then, we can still use the WOP \eqref{eq:wf}, although the value of $W$ will depend on the choice of $\Tau_1,\Tau_2$. Suppose the main goal is to explore the transition to synchrony. In that case, we suggest fixing $\Tau_1,\Tau_2$ and changing a parameter (e.g., coupling strength) to see a difference in WOP between synchronous and asynchronous regimes. All the arguments above assume large values of $N$; 
for small population sizes $N$, the finite-size effects are rather prominent so that the very notion of synchrony becomes fuzzy. We will illustrate this in Section~\ref{sec:ar}. 

\subsection{Irregular mean fields}
\label{sec:imf}

The described approach assumes that the mean fields in the thermodynamic limit are regular (periodic or quasiperiodic) and relies on identifying the corresponding regular component in the observation by virtue of the ACF analysis. Here, we discuss the case of irregular mean fields that occur if a 
solution of the mean-field equations (\ref{eq:op-1}-\ref{eq:op-3})  in the thermodynamic limit is chaotic. Examples from the literature include globally coupled Stuart-Landau oscillators~\cite{hakim1992dynamics,nakagawa1994collective,chabanol1997collective,ku2015dynamical,clusella2019between,leon2022enlarged} and coupled neurons~\cite{olmi2011collective,pazo2016quasiperiodic,ratas2018macroscopic}. 

We argue that the presented technique can still be used if the mean fields are weakly irregular. 
The idea is basically the same as in the case of finite-size fluctuations (Section~\ref{sec:fse}).
Indeed, the ACF eventually decreases to zero for an irregular process, so its power spectrum does not contain discrete components, and the WOP \eqref{eq:wf} vanishes. However, suppose the process is close to a regular one, with narrow peaks in the power spectrum. In that case, the ACF decays rather slowly at large times (the characteristic time is the diffusion time of the phase of the nearly-periodic oscillations). As a result, the calculation of the WOP according to Eq.~(\ref{eq:wf1}) provides a finite value. If one chooses $\Tau_1$ in Eq.~(\ref{eq:wf1}) to be larger than the time of an initial drop of the ACF (due to a possible broadband component of the power spectrum), and $\Tau_2$ in Eq.~\eqref{eq:wf1} is smaller than the characteristic decay time of a long-living component of the ACF, then expression \eqref{eq:wf1} will still provide a reasonable estimate of the intensity of the nearly regular component of the mean field variations. 

If the irregularity of the mean fields is large, and there is no clear time scale separation between different epochs of the ACF decay, the application of the ACF analysis seemingly brings no advantage compared to the calculation of the variance of the observed mean field.  

\subsection{Non-identical units}
\label{sec:niu}

Above we concentrated on the case of identical globally coupled oscillators; here we shortly argue, that the approach can also be applied to non-identical units, provided they are irregular enough.
A typical way to introduce inhomogeneity in the population is to assume that the dynamics \eqref{eq:geq-1} depends additionally on a parameter $b_k$, and this parameter is different for different oscillators. In the thermodynamic limit, one describes the inhomogeneity with a distribution $g(b)$. As a result, the probability density contains $b$ as an additional variable, $w(\vec{x},t|b)$ and averaging in \eqref{eq:op-3} contains additional averaging over the distribution of $b$. With such a modification, the basic conclusions of Sections \ref{sec:gmfc}-\ref{sec:elim} are still valid. Namely, at synchrony the mean fields oscillate regulary, while at asynchrony they are constants. Each unit has a regular and irregular component, and the former can be identified via the ACF analysis. The only modification is that now sampled observations (Section \ref{sec:cpo}) are not sample-independent, because the units are not identical. As a result,
in Eq.~\eqref{eq:acf2}, both components will depend on the sample $\mathcal{S}_M$. Nevertheless, the separation of time scales will be still valid, and the Wiener's formula \eqref{eq:wf} will provide
the intensity of the regular component. This consideration shows that the WOP will now depend on the sample $\mathcal{S}_M$. In particular, we expect to have different values if just one unit is observed. Nevertheless, the contrast between the asynchronous and synchronous states will be still present for any sample $\mathcal{S}_M$.

Below, we apply the described approach to several examples. 
One of the main questions is: how well can we characterize synchrony with an observation of just one unit?
We will start with strongly irregular units in Sections~\ref{sec:ar},\ref{sec:ca}.
In Section~\ref{sec:soq}, we will explore identical noisy limit-cycle oscillators, also in the case of a weak noise.
In the last example (Section~\ref{sec:kur}) we will apply our 
approach to the Kuramoto model of non-identical phase oscillators.

\section{Noise-induced oscillations}
\label{sec:ar}
A recent paper~\cite{klinshov2021noise} studied the dynamics of globally coupled active rotators. 
The equations for the units described by the phase-like variables $\phi$ 
are formulated as
\begin{equation}
\dot\phi_k=1-b\sin\phi_k+\frac{\e}{N}\sum_j\sin(\phi_j-\phi_k)+\sqrt{2\sigma^2}\xi_k(t)\;.
\label{eq:ar}
\end{equation}
With the choice $b>1$, adopted in Ref.~\onlinecite{klinshov2021noise}, the uncoupled noise-free systems have a stable steady state and do not oscillate. In the presence of noise, the dynamics of uncoupled units are purely noise-induced and thus highly irregular. 
The coupling term $\sim \e$
is of Kuramoto type; it can be written in terms of mean fields 
\begin{equation}
C(t)=\frac{1}{N}\sum_k \cos\phi_k(t),\qquad S(t)=\frac{1}{N}\sum_k \sin\phi_k(t)\;.
\label{eq:armf}
\end{equation}
In the thermodynamic limit, the dynamics of the probability density $w(\phi,t)$ reduces to a nonlinear Fokker-Planck equation of type \eqref{eq:op-1}. The regimes in this nonlinear PDE have been carefully studied in  Ref.~\onlinecite{klinshov2021noise}, and a domain of parameters where the density  $w(\phi,t)$ varies periodically in time, has been identified (together with other domains, where the attractor in the nonlinear PDE is a steady state). 

Below, we fix the parameters $\e=0.6$ and $b=1.025$. For these values, the phase diagram elaborated in  Ref.~\onlinecite{klinshov2021noise}
predicts a time-periodic (synchronous) solution for $\sigma^2=0.04$ and a stationary (asynchronous) solution at $\sigma^2=0.06$.

\subsection{Large number of active rotators}
\label{sec:arln}

\begin{figure}[!htb]
\centering
\includegraphics[width=\columnwidth]{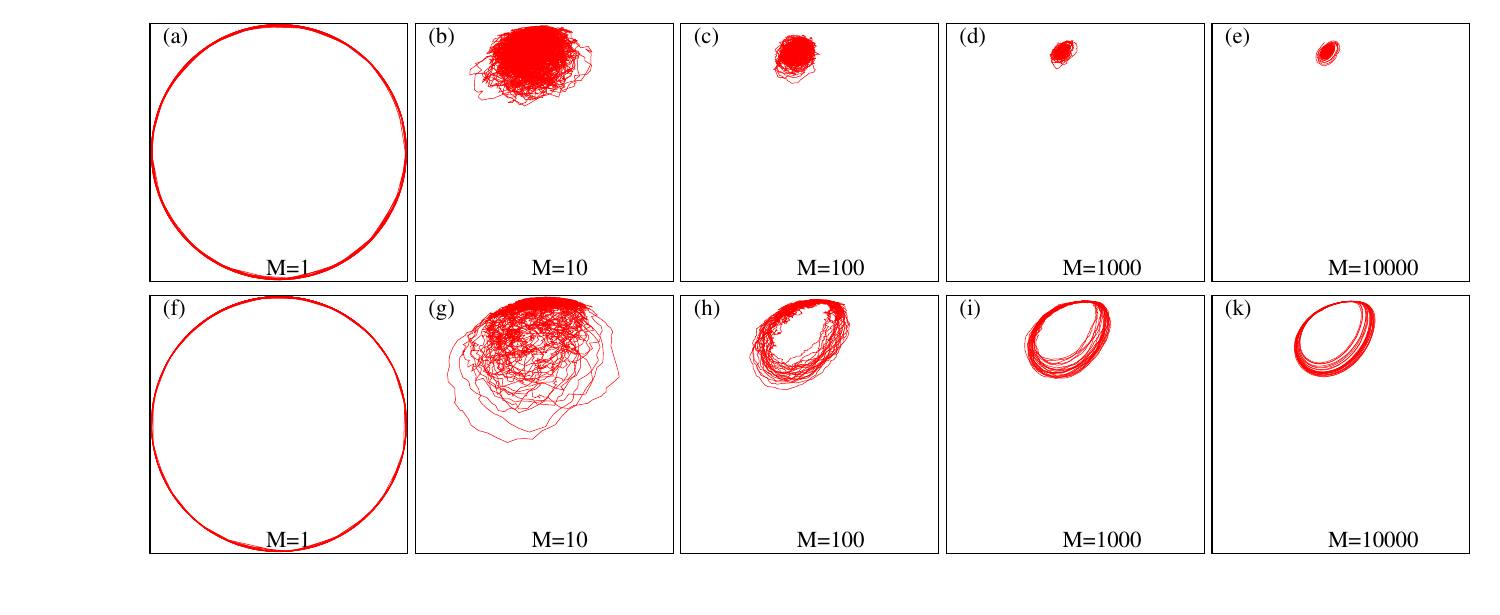}
\caption{Synchronous and asynchronous states of noise-driven active rotators \eqref{eq:ar}. Presented are 
the  ``phase planes'' of observables $C_M(t),S_M(t)$, see Eq.~\eqref{eq:armf}, 
for different sample sizes $M$ (the total size of the ensemble is $N=10^4$). 
Upper row (panels (a-e)):  $\sigma^2=0.06$ (asynchrony), bottom row (panels (f-k)): $\sigma^2=0.04$ (synchrony). The scales in all panels are the same, $-1\leq C_M,S_M\leq 1$.}
\label{fig:ar-2}
\end{figure}

Figure~\ref{fig:ar-2} illustrates how these solutions for an ensemble of $N=10^4$ oscillators
are represented by finite samples.
Here, we depict  mean fields $C_M$, $S_M$ calculated following Eqs.~(\ref{eq:armf},\ref{eq:pmf}) over  randomly chosen subsets of $M$ units, starting with $M=1$ (just one active rotator is observed)
up to $M=N=10^4$, what corresponds to full mean fields \eqref{eq:armf}. In the projections on the partial mean field phase planes $(C_M,S_M)$
shown in Fig.~\ref{fig:ar-2}, one can see for $M\geq 100$ and $\sigma^2=0.04$ a ``topological limit cycle'', where the trajectory rotates with a void in the center. In contradistinction, 
for $M= 10$ and $\sigma^2=0.04$, and all $M$ and $\sigma^2=0.06$ one does not see a topological limit cycle (such a cycle is, of course, present for $M=1$ because the observables defined in Eq.~(\ref{eq:armf}) fulfill $C_1^2+S_1^2=1$). We note here that the existence of a void in a two-dimensional projection of the mean fields has been suggested in Ref.~\onlinecite{Temirbayev_etal-12} as a practical criterion for synchrony in small populations (cf. a more detailed analysis of this criterion for the Kuramoto model in Ref.~\onlinecite{Peter-Pikovsky-18}).

\begin{figure}[!htb]
\centering
\includegraphics[width=\columnwidth]{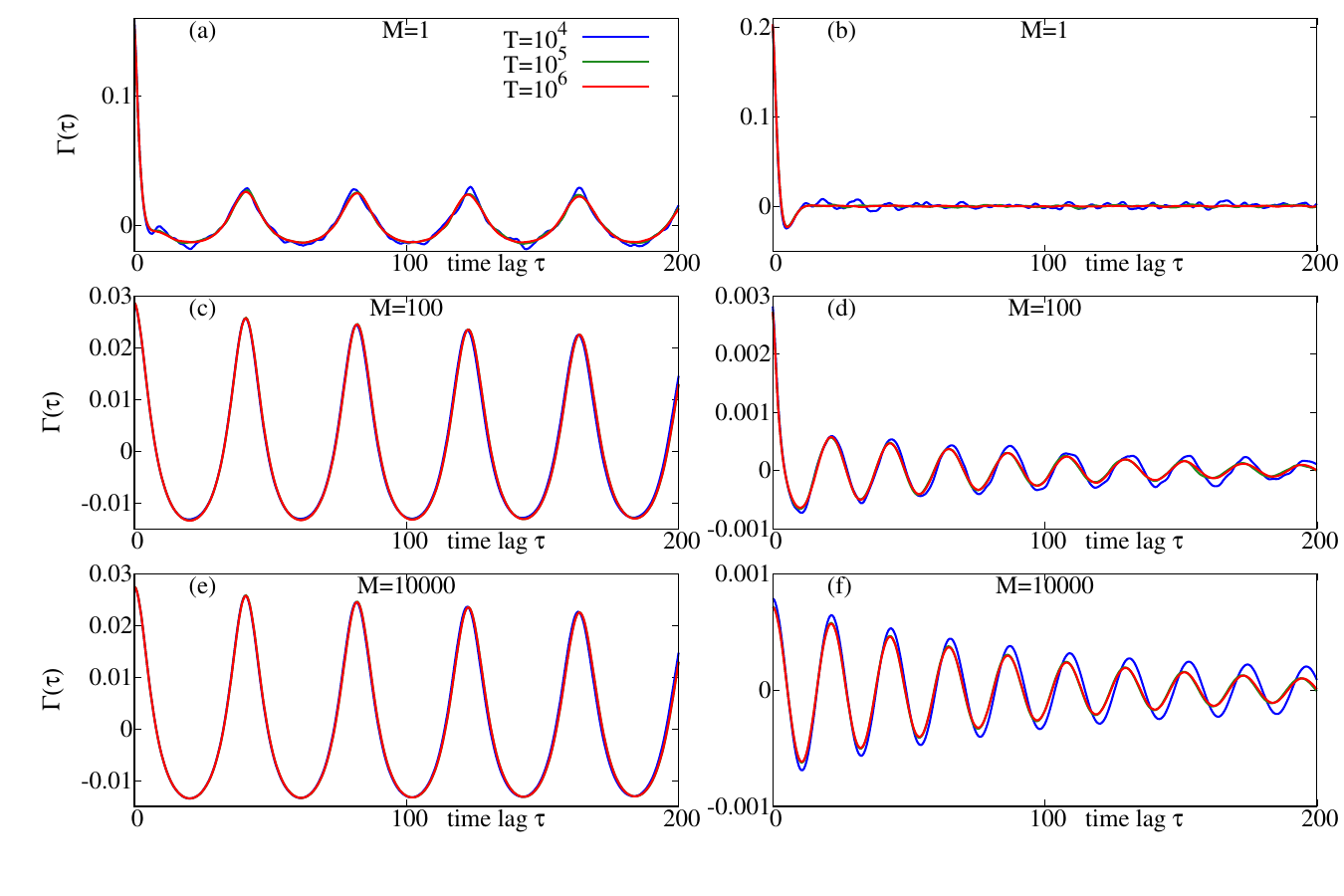}
\caption{Autocorrelation functions of different sampled observables $C_M(t)$, calculated using different time intervals $T$. Left column (panels (a,c,e)): small noise $\sigma^2=0.04$; right column (panels (b,d,f)): large noise $\sigma^2=0.06$. Notice that the vertical scales are different
in all panels.}
\label{fig:ar-3}
\end{figure}

The autocorrelation functions of observables $C_M(t)$ calculated according to Eq.~\eqref{eq:acf} 
are shown in Fig.~\ref{fig:ar-3}. Here we present only small time lags $0\leq\tau<200$, to illustrate 
convergence with the averaging time $T$. One can see that  all the autocorrelation functions $\Gamma_M(t)$ averaged over $T=10^5$ and $T=10^6$ practically coincide, while some deviations are seen for the short averaging time $T=10^4$.

\begin{figure}[!htb]
\centering
\includegraphics[width=0.9\columnwidth]{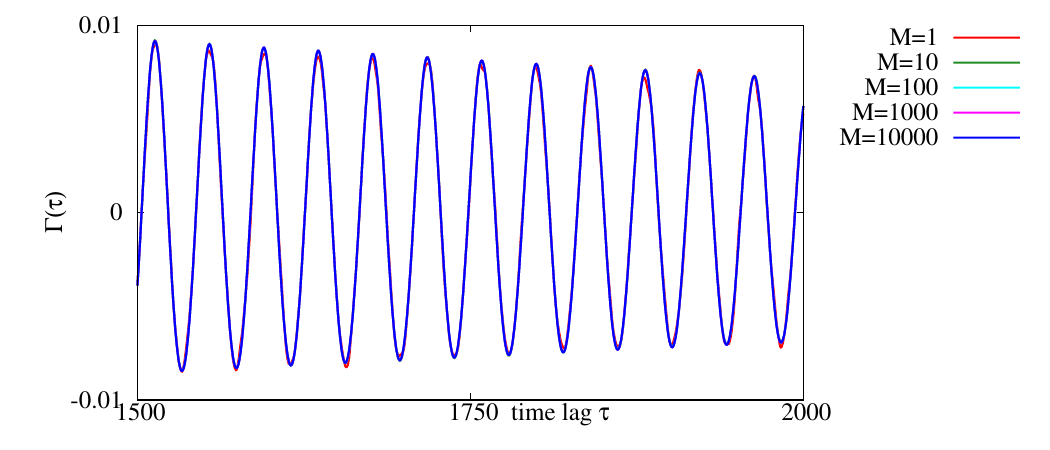}
\caption{The ACFs for the synchronous case $\sigma^2=0.04$, $N=10^4$, for different sample sizes $M$ at large time lags. The averaging time is $T=10^6$.
All the curves with $10\leq M\leq 10^4$ practically coincide.}
\label{fig:ar-4}
\end{figure}

\begin{figure}[!htb]
\centering
\includegraphics[width=0.9\columnwidth]{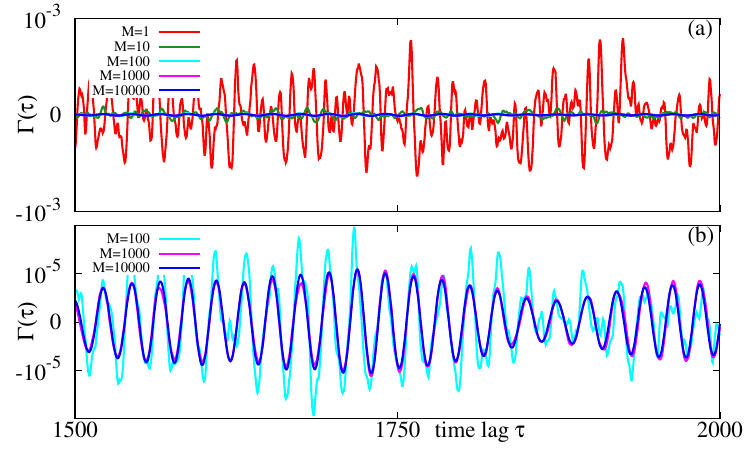}
\caption{The ACFs for the asynchronous case $\sigma^2=0.06$, $N=10^4$,  for different  sample sizes $M$ at large time lags.  The averaging time is $T=10^6$. Panel (b) shows the ACFs for $M\geq 100$ in the appropriate scale.}
\label{fig:ar-5}
\end{figure}

Our primary interest is in the behavior of the ACFs at large time lags; we show these data in Figs.~\ref{fig:ar-4},\ref{fig:ar-5} for $1500\leq \tau\leq 2000$. In the synchronous case (Fig.~\ref{fig:ar-4}), the ACFs for different
sample sizes $M$ practically coincide if the averaging time $T$ is large. This confirms the concept of section~\ref{sec:gc} that the regularity of the mean fields can be extracted from samples with any $M$, even with $M=1$. Another observation is that the amplitude of the ACF slowly decreases. The reason for this are finite-size fluctuations for $N=10^4$, as discussed in Section~\ref{sec:fse}.

The ACFs in the asynchronous case (Fig.~\ref{fig:ar-5}) have relatively small values and 
demonstrate strong dependence on the sample size $M$ (thus, we show two panels, omitting cases $M=1,10$ in panel (b)). A remarkable observation is that at a presented time interval of averaging $T=10^6$, the correlations of the fields with $M=1000$ and $M=N=10^4$ are rather regular, although small in amplitude. This issue needs further investigation.

\begin{figure}[!htb]
\centering
\includegraphics[width=0.9\columnwidth]{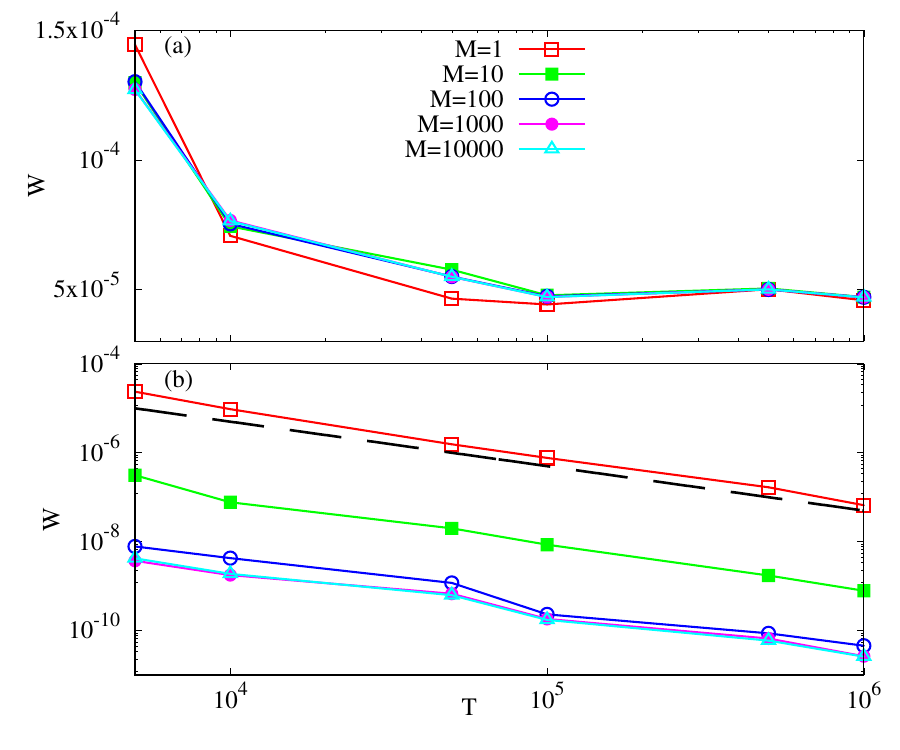}
\caption{The Wiener order parameter $W$ for different sample sizes $M$ and ACF averaging times $T$
(other parameters: $\Tau_1=750$, $\Tau_2=2000$).
Panel (a): synchronous case $\sigma^2=0.04$, $N=10^4$. Note that the vertical scale is linear.
Panel (b): asynchronous case $\sigma=0.06$, $N=10^4$; the verical scale is logarithmic. The black dashed line shows the scaling $W\sim T^{-1}$. }
\label{fig:ar-6}
\end{figure}

Finally, we compare the WOPs $W$ for the two cases. In Fig.~\ref{fig:ar-6}, we show $W$ for different sample sizes $M$ and different averaging time intervals $T$. As is expected from the discussion in
Section~\ref{sec:elim}, dependence on $T$ is obviously different for the two cases. 
In the synchronous case (panel (a)), $W$ saturates at some value for $T\geq 10^5$, so this value can be taken as a correct estimate of $W$. Furthermore, this value is practically the same for all $M$, as the coincidence of the ACFs in Fig.~\ref{fig:ar-4} suggests. On the contrary, for the asynchronous case, the values of $W$ decay $\sim T^{-1}$ at large $T$. While for purely fluctuating samples 
one would also expect $W\sim M^{-2}$; such scaling is valid only for $M=1,10$; for $M\geq 100$, apparently, this scaling is not applicable due to the relative regularity of fluctuations of ACFs.

\subsection{Finite size effects for small ensembles of active rotators}
\label{sec:arsn}

In Section~\ref{sec:arln} above, we considered a rather large ensemble of irregular oscillators  with $N=10^4$. 
Here, we explore, for the same system, how the discussed methods of synchrony characterization work for small population sizes.
In particular, we explore cases $N=10,\;100,\;1000$.

\begin{figure}[!htb]
\centering
\includegraphics[width=\columnwidth]{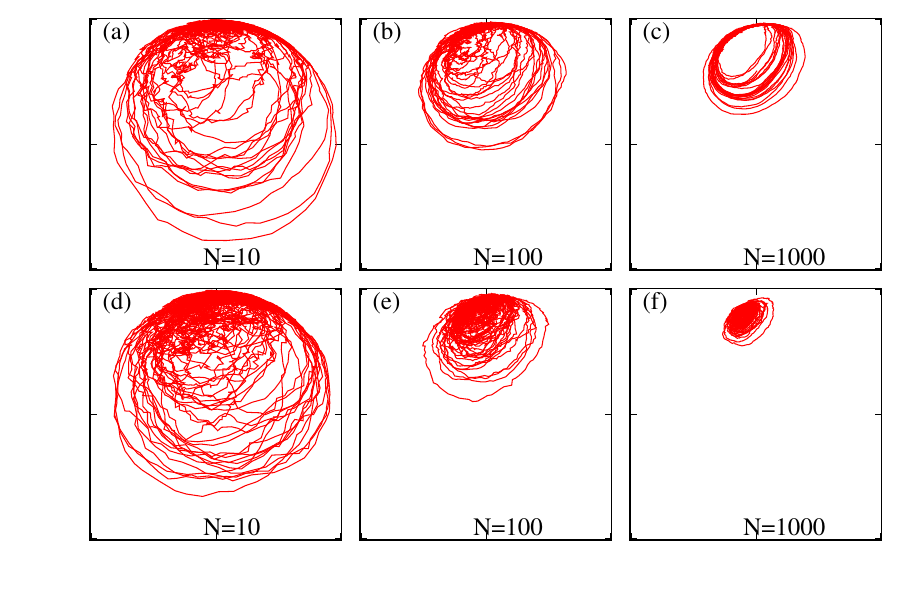}
\caption{Phase portraits of the complete mean fields $C_N$ vs $S_N$ for small system sizes $N=10,\;100,\;1000$ and two values
of the noise level.}
\label{fig:arf-1}
\end{figure}

First, in Fig.~\ref{fig:arf-1} we present the phase portraits $(C_N(t),S_N(t))$ for the small populations, for $\sigma^2=0.04$ and $\sigma^2=0.06$. One can see a clear difference between two noise values for $N=1000$, while for $N=10,100$ for both noise levels, there
is no void in the phase portraits around the center of rotations. This allows for a preliminary conclusion that synchrony detection with other methods will possibly work for $N=1000$, but might be hard for  $N=10,100$.

\begin{figure}[!htb]
\centering
\includegraphics[width=0.9\columnwidth]{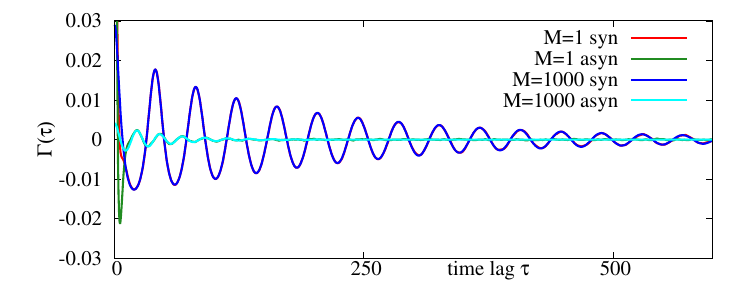}
\caption{Characterization of the regimes depicted in Fig.~\ref{fig:arf-1} (c,f), $N=1000$, with the ACFs.
Both for small noise (synchrony in the thermodynamic limit), $\sigma^2=0.04$, and for large noise (asynchrony in the thermodynamic limit), $\sigma^2=0.06$, the ACFs from one element and from the complete mean field coincide after a small initial interval of time lags.}
\label{fig:arf-2}
\end{figure}

\begin{figure}[!htb]
\centering
\includegraphics[width=0.9\columnwidth]{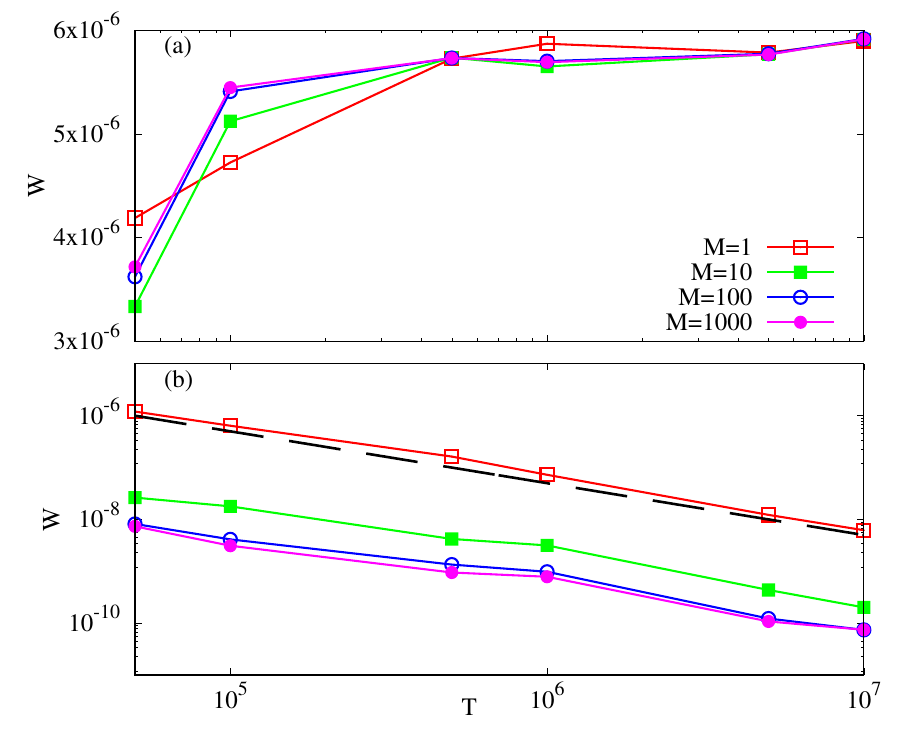}
\caption{The same as Fig.~\ref{fig:ar-6}, but for the case $N=1000$. One can see that
the WOP reasonably contrasts synchronous and asynchronous states despite pronounced finite-size irregularity. Parameters of the WOP calculation: $\Tau_1=200$, $\Tau_2=600$.
}
\label{fig:arf-3}
\end{figure}

In Fig.~\ref{fig:arf-2}, we show different autocorrelation functions for $N=1000$. 
First, we notice that the ACFs for $M=1$ and $M=1000$ practically coincide except for a region of very small time lags (one cannot see the red and the green curves in Fig.~\ref{fig:arf-2} because these curves are overlapped with the blue and the cyan ones). Altogether, the different time behavior of the ACFs delivers a strong contrast between synchrony and asynchrony for $N=1000$ (because the blue curve for $\sigma^2=0.04$ demonstrates a much more ordered oscillatory tail compared to the cyan curve for $\sigma^2=0.06$). On the other hand, since even in the synchronous case, the ACF significantly decays after 10 periods, for the calculation of the WOP $W$ we have to choose relatively small values $\Tau_1,\Tau_2$. With such a choice,
this order parameter allows for a reasonable synchrony characterization, as illustrated in Fig.~\ref{fig:arf-3}. 

Performing the same ACF analysis for small population sizes,
we found that for $N=100$ the difference in the correlation lengths for the two values on noise strength is rather small, and for $N=10$ there is practically no difference.  
These results confirm a general expectation in Section~\ref{sec:gc} that a transition to synchrony for small population sizes becomes fuzzy.

\section{Chaotic oscillators}
\label{sec:ca}

\begin{figure}[!htb]
\centering
\includegraphics[width=0.9\columnwidth]{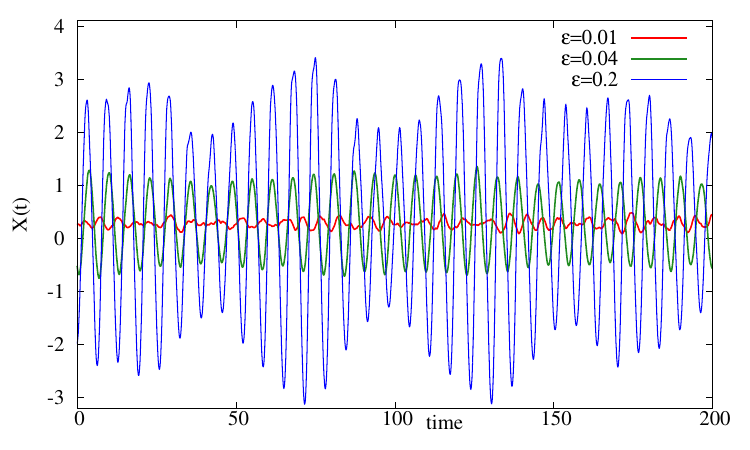}
\caption{Time series of the global mean field $X=N^{-1}\sum_k x_k(t)$ for coupled chaotic
R\"ossler oscillators \eqref{eq:roes} for different values
of the coupling strength $\e$. The ensemble size is $N=10^4$.}
\label{fig:fr-1}
\end{figure}

Here we describe synchronous states in coupled R\"ossler oscillators, first reported in Ref.~\onlinecite{Pikovsky-Rosenblum-Kurths-96}:
\begin{equation}
\begin{aligned}
\dot x_k&=-y_k-z_k+\frac{\e}{N}\sum_j x_j\;,\\
\dot y_k&=x_k+ay_k\;,\\
\dot z_k&=b+(x_k-c)z_k\;.
\end{aligned}
\label{eq:roes}
\end{equation}
We use parameters $a=0.25$, $b=0.4$, $c=8.5$, at which each R\"ossler system has a funnel-type attractor, i.e., the oscillators have an ill-defined phase (cf. phase portraits with $M=1$ in Fig.~\ref{fig:fr-2} below).
For this system we use the observables 
\[
X_M(t)=\frac{1}{M}\sum_{k\in\mathcal{S}_M} x_k(t)\;,\qquad Y_M=\frac{1}{M}\sum_{k\in\mathcal{S}_M} y_k(t)\;.
\]

With the increase of the coupling strength $\e$, first periodic oscillations of the mean field appear, then the amplitude of these oscillations grows, and they become modulated. We illustrate this in Fig~\ref{fig:fr-1}, where we show the time series  
of the observable $X_N(t)$ for $N=10^4$ and different values of $\e$. For $\e=0.01$, the mean field fluctuates around a constant without any visible regularity; we attribute the fluctuations to finite-size effects and denote this state as an asynchronous one. In contrast, we observe regular macroscopic variations of the mean field for $\e=0.04$ and $\e=0.2$ and correspondingly denote these states as synchronous. 

\begin{figure}[!htb]
\centering
\includegraphics[width=\columnwidth]{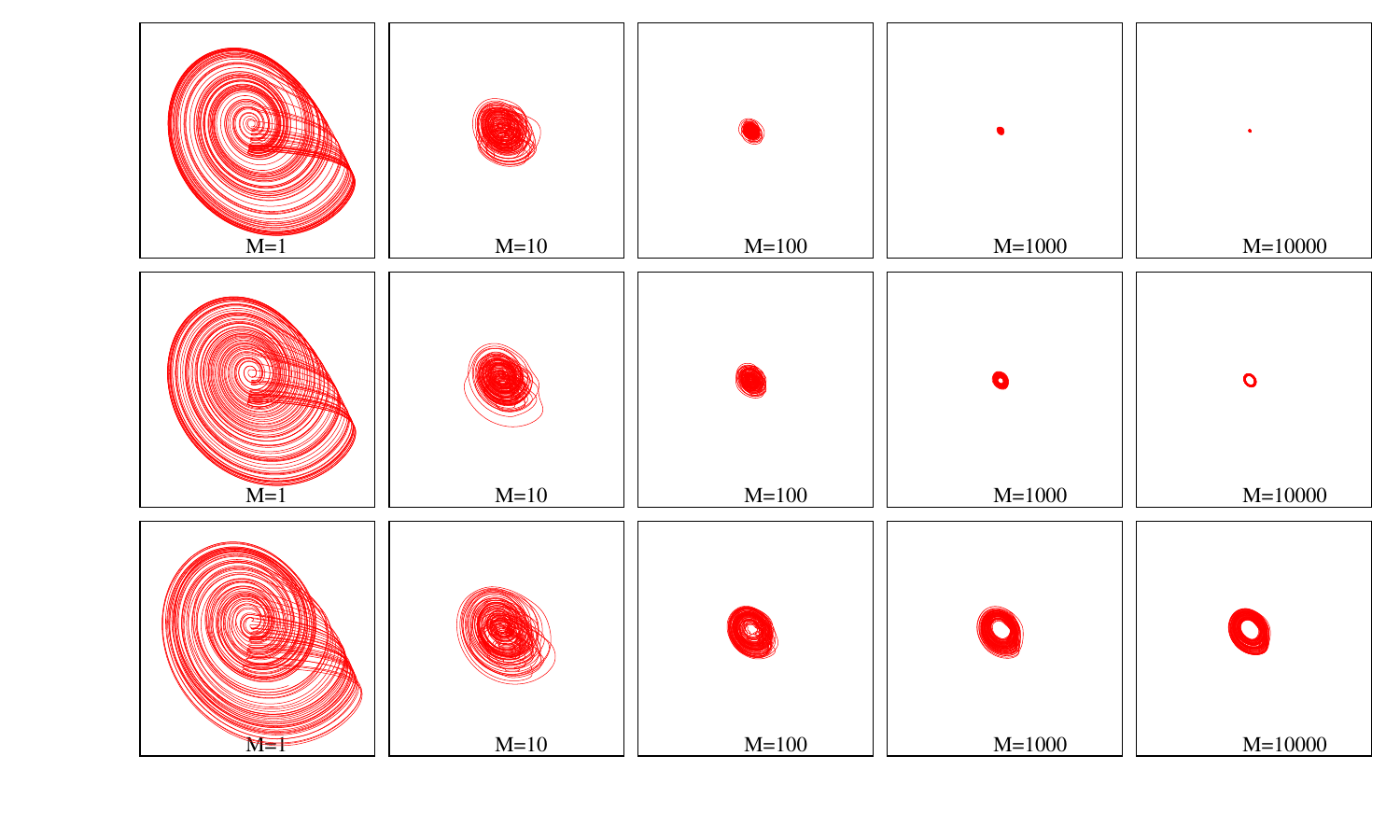}
\caption{Phase portraits $(X_M,Y_M)$ for $M=1,10,100,10^3,10^4$,
for synchronous and asynchronous regimes (from top to bottom: $\e=0.01$, $\e=0.04$, $\e=0.2$).}
\label{fig:fr-2}
\end{figure}

In Fig.~\ref{fig:fr-2} we show trajectories of partial mean fields $(X_M,Y_M)$ for $M=1,10,100,10^3,10^4$,
for synchronous and asynchronous regimes. 
One can clearly see that these observables show very similar patterns for $M=1$ and $M=10$ for all $\e$. At $M=100$ one can recognize a regularity in the partial mean fields for  $\e=0.2$, while a regularity for $\e=0.04$ is only visible for $M\geq 10^3$. 

\begin{figure}[!htb]
\centering
\includegraphics[width=\columnwidth]{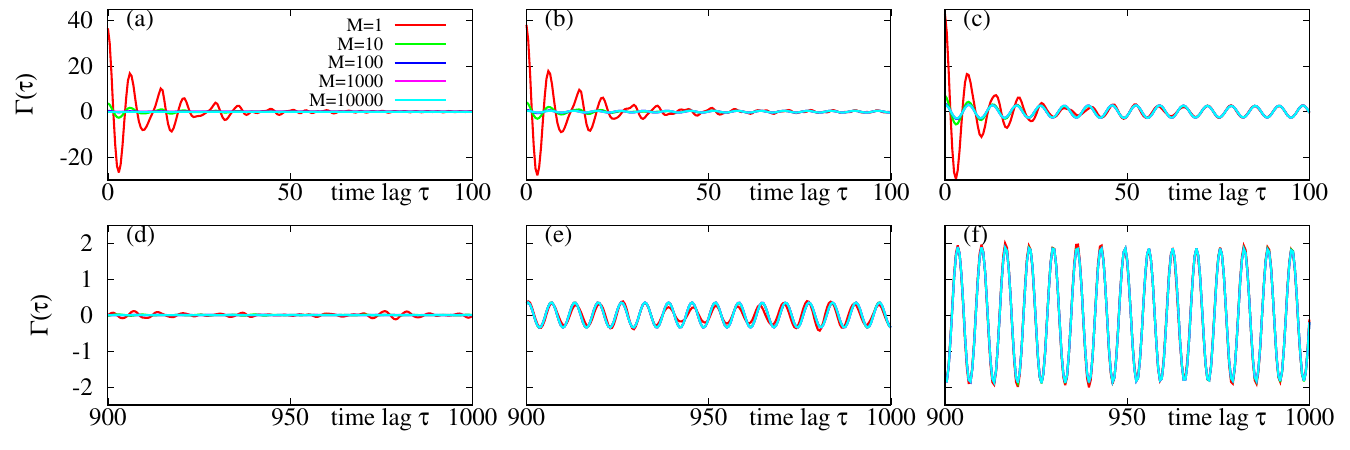}
\caption{Autocorrelation functions at small time lags (panels (a-c)) 
and at large time lags (panels (d-f)) 
for different $M$ and $\e$ (averaging time $T=10^6$). Panels (a,d): asynchronous state $\e=0.01$; here, all the autocorrelation functions at large
time lags fluctuate around zero. Panels (b,e): $\e=0.04$; panels (c,f): $\e=0.2$. Here, the ACFs for observables with different $M$ nearly coincide at large time lags, where they demonstrate periodic behavior.}
\label{fig:fr-3}
\end{figure}

Finally, we apply the ACF analysis to reveal synchrony from observations with small sample sizes $M$.
In Fig.~\ref{fig:fr-3} we  show the ACFs $\Gamma [X_M](t)$ for the observables $X_M$, for all the three selected values
of the coupling strength $\e$. 
For $\e=0.01$ (panels (a,d)), the ACFs for all $M$ fluctuate at a small level for large time lags, confirming the absence of synchrony.
For the synchronous cases $\e=0.04$ (panels (b,e)) and $\e=0.2$ (panels (c,f)), at large time lags the autocorrelation functions for all $1\leq M\leq N$ nearly coincide. This supports the conclusion that observation of just one unit in a large population allows for revealing synchrony by virtue of our method.

\section{Noisy limit-cycle oscillators}
\label{sec:soq}
In this section, we present an example where we vary the irregularity level at each unit to see the impact of this level on the performance of the method.
We take the system of $N$ Stuart-Landau oscillators with global nonlinear coupling and add independent Gaussian white noise terms $\sim\sigma$ to each unit:
\begin{equation}
\dot a_k=(1+5\ii)a_k-|a_k|^2a_k+\e A-\e_{nl}|A|^2A+\sigma\xi_k(t)\;.
\label{eq:SL1}
\end{equation}
Here $a_k$ are complex variables, $k=1,2,\ldots,N$, and $A=N^{-1}\sum_{k=1}^N a_k$ is the complex mean field. The terms 
$\xi_k(t)$ are Gaussian white noises with zero mean and unit intensity~\footnote{Notice that the normalization of the noise strength differs from that used in Eq.~(\ref{eq:ar})}. 
Analysis of this system performed in Ref.~\onlinecite{Rosenblum-Pikovsky-15} without noisy perturbations shows that for specific parameter values the ensemble exhibits partial synchronization, such that the mean field is periodic, but its frequency differs from that of the oscillators and the latter are thus quasiperiodic. 

\subsection{Linear coupling} 
\begin{figure}[!htb]
\centering
\includegraphics[width=0.9\columnwidth]{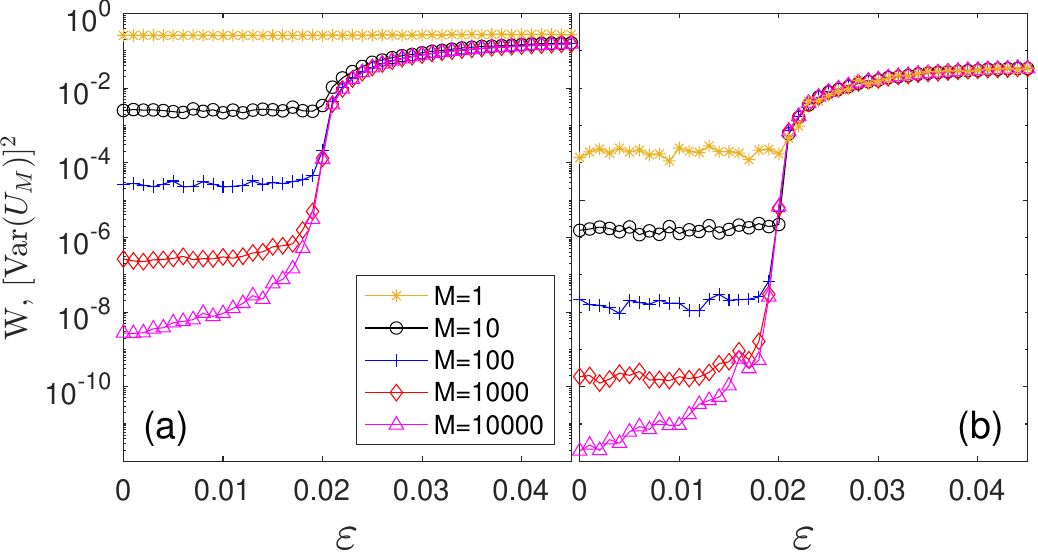}
\caption{The variance measure of synchrony (a) and the Wiener order parameter (b) 
of the ensemble of noisy Stuart-Landau oscillators~(\ref{eq:SL1}) with global linear coupling
as functions of the coupling strength $\varepsilon$. Different curves in each panel correspond to
observation of $M=1$, $M=10,\ldots$, $M=N=10^4$ oscillators.  
Notice that $W$ successfully reveals the synchronization transition even from an observation of a single oscillator. Next, for the $M>1$ cases, the quantification with the WOP provides an about two orders of magnitude stronger contrast between the states before and after transition.  
}
\label{fig:sl-1}
\end{figure}

Here, we fix $\e_{nl}=0$, $\sigma=0.2$, and vary real parameter $\e$ to trace the synchronization transition. For this purpose, we exploit two traditional measures (the mean-field variance and the Kuramoto order parameter), as well as the WOP. We simulate the ensemble of  $N=10^4$ units and take 
$U_M=M^{-1}\text{Re}\big(\sum_{k=1}^M a_k\big)$,   $M\le N$, as our observable. (We use $10^6$ time points sampled with the step $0.05$ to compute the autocorrelation function and fix the values 
$\Tau_1=1.25\cdot 10^4$ and $\Tau_2=2\cdot 10^4$ for the computation of $W$.)
The results are illustrated in Fig.~\ref {fig:sl-1}. Since $W$ is proportional to the squared variance of the process, we plot $[\text{Var}(U_M)]^2$ in  the panel (a) and $W$ in the panel (b). (We do not show the KOP since its variation is, up to a nearly constant factor, very similar to that of the variance measure.)  

The main outcome is that the Wiener order parameter successfully traces the synchronization transition from an observation of a single oscillator, while the variance measure certainly fails here.  Furthermore, for partial observations with $M>1$, when both measures work, the WOP is more efficient: the contrast between the asynchronous and synchronous states is much stronger.  

\subsection{Nonlinear coupling} 
Here, we fix $\e=3+0.3\ii$ and $\e_{nl}=8$. With this nonlinear coupling, the noise-free ensemble 
in the thermodynamic limit $N\to\infty$ exhibits the harmonic mean-field solution, while individual units remain not locked to the field and thus dynamics of each unit is quasiperiodic. This quasiperiodicity will be naturally seen in the one-unit sampling $M=1$, but will be smeared for larger $M$. 

\begin{figure}[!htb]
\centering
\includegraphics[width=0.9\columnwidth]{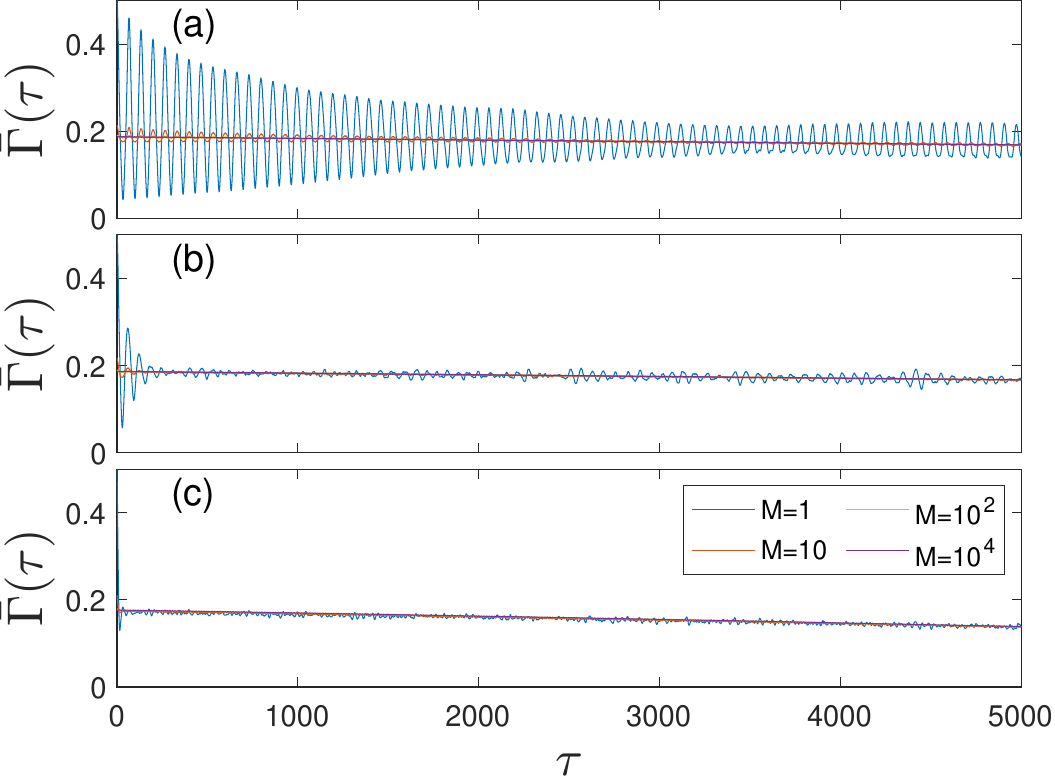}
\caption{The upper envelopes of the correlation functions for the nonlinearly coupled ensemble of Stuart-Landau oscillators~(\ref{eq:SL1}), for different values of the noise intensity: $\sigma=0.02$ (a), $\sigma=0.1$ (b), and $\sigma=0.5$ (c) and different values of $M$.
}
\label{fig:sl-2}
\end{figure}

Figure~\ref{fig:sl-2} demonstrates the simulation 
results for $N=10^4$ and different values of the noise intensity. Here, for a better visibility,  we plot 
the upper envelopes of the autocorrelation functions; thus, a nearly constant value of $\bar{\Gamma}$ at large $\tau$ indicates for a purely periodic correlation function with the corresponding amplitude of oscillations. 

The main conclusion is that regularity of the oscillators is an obstacle for observations with $M=1$: for a weak noise, $\sigma=0.02$ (panel (a)), the ACF envelope for $M=1$ strongly oscillates, indicating modulation of the ACF, as expected for a signal that is close to a quasiperiodic one. Correspondingly, computation of $W$ from one oscillator will provide the sum of intensities of periodic components on the oscillator's frequency and
that of the mean field (and their harmonics). For this level of noise, oscillations of the ACF envelope can be also seen for $M=10$,
but they are much smaller. So, due to relative regularity, the correlation decay is very slow; hence, inference of synchrony from observation of one or several units requires an extremely long observation. 

The situation changes for an intermediate noise, $\sigma=0.1$ (panel (b)). Now, we see that the oscillation of the ACF envelope for $M=10$ decays quickly so that the partial mean field $U_{10}(t)$ suffices for the computation of $W$. For stronger noise, $\sigma=0.5$
(panel (c)), even observation of one oscillator (after a short decay time) yields the same ACF as the complete observation of the mean field.   This confirms the overall estimation that the proposed method should work better for irregular oscillators and might not be optimal when local units are highly regular.

Two remarks are in order. First, we plotted only the envelopes of the ACFs, but we checked that when they coincided, the ACFs coincided as well. Second, in panel (c), we see that the envelope is not constant but slightly decays so that, strictly speaking, the ACFs are not stationary 
for the presented range of $\tau$. This decay is because the mean field is not truly periodic due to the large though finite ensemble size, as discussed in Section~\ref{sec:fse}.    

\subsection{Nonidentical noisy oscillators}

\begin{figure}[!htb]
\centering
\includegraphics[width=0.9\columnwidth]{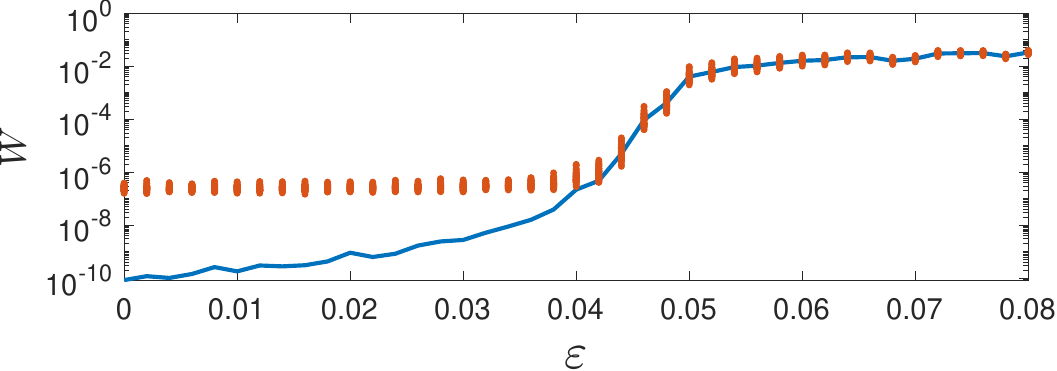}
\caption{The Wiener order parameter $W$ as a function of the coupling parameter $\e$ for the 
system of $N=10^4$ inhomegeneous noisy oscillators. Blue line shows the values obtained from the mean field of the ensemble. Red dots present the results of partial observations, i.e., the partial mean field is computed from 20 randomly chosen units. For each value of $\e$, we perform 50 trials
which provide generally different results.
Since the units are non-identical, the order parameter computed from a partial observation depends on the chosen subset of units, what explain the vertical scattering of the points. 
Notice that the scattering decreases with $\e$ because of the tendency to synchrony.
}
\label{fig:sl-3}
\end{figure}

To illustrate ideas of Section~\ref{sec:niu} about the approach's applicability to non-identical units, we take a large ensemble of $N=10^4$ slightly inhomogeneous noisy oscillators. 
The model now  reads:
\begin{equation}
\dot a_k=(1+\ii\w_k)a_k-|a_k|^2a_k+\e A+\sigma\xi_k\;,
\label{eq:SL2}
\end{equation}
where frequencies $\w_k$ are taken from a Gaussian distribution with the mean value $\bar\w=5$ and 
standard deviation $0.02$; the noise strength is $\sigma=0.2$, and the coupling strength is varied.
As before, we use $U_M=M^{-1}\text{Re}\big(\sum_{k=1}^M a_k\big)$ to 
compute the WOP $W$ as a function of $\e$.
 Figure~\ref{fig:sl-3} demonstrates the synchronization transition revealed from the global mean field, $M=N=10^4$, and observations of a small subpopulation, $M=20$. Since the units are non-identical, the results definitely depend on the chosen subset of units. To check that this dependence is weak, we performed calculations with $50$ random samples of $M=20$ units and presented the results for all these samples as red dots. We see that all the samples give very similar results. Thus, an observation of only 2\% of the oscillators reliably reveals the transition to synchrony.  

\section{Kuramoto model}
\label{sec:kur}
Here we apply our approach to the standard Kuramoto model of globally coupled phase oscillators
\begin{equation}
\dot\vp_k=\w_k+\frac{\e}{N}\sum_{j=1}^N \sin(\vp_j-\vp_k),\quad k=1,2,\ldots,N\;.
\label{eq:kur}
\end{equation}
The main differences to the cases above are: (i) the specific units, if uncoupled, are not irregular (chaotic or noisy), but regular periodic oscillators;
(ii) the units are not identical but differ in natural frequencies $\w_k$. The latter property ensures the existence of an asynchrony state for small coupling strengths $\e$. 

In simulations below we take a Gaussian distribution of frequencies $N(\bar{\w},1)$ with unit variance centered at frequency $\bar{\w}=2\pi$. The theory~\cite{Peter-Pikovsky-18} predicts in the thermodynamic limit a transition to periodic global oscillations at $\e_{cr}=2\sqrt{2/\pi}\approx 1.596$; the dependence of the Kuramoto order parameter (KOP) $R=|\langle e^{\ii\vp}\rangle|$ on $\e$ can be represented in the parametric form with real-valued parameter $p>0$ as 
\begin{equation}
R=e^{-p}\sqrt{\pi p/2}[I_0(p)+I_1(p)],\quad \e=2\sqrt{p}/R\;,
\label{eq:kop}
\end{equation}
where $I_0,I_1$ are modified Bessel function of the first kind. 

Below, we consider finite ensembles with moderate numbers of units $N$. We will use regular sampling of natural frequencies $\w_k$,
like in Refs.~\onlinecite{carlu2018origin,Peter-Pikovsky-18}, calculating them from the corresponding quantiles of the normal distribution. Because the analytic solution \eqref{eq:kop} in the thermodynamic limit  is available, we will compare the numerical
results with the theory. For this, we need to relate the KOP to the WOP, see 
Eqs.~(\ref{eq:wf},\ref{eq:wf1}). We will use the local observable
$u(\vp)=\cos\vp$ for the latter. Thus, the fully observed  (i.e., calculated for the full ensemble size $N$)  mean field $U(t)=\text{Re}(Z(t))$ is the real part of the complex Kuramoto order parameter $Z(t)=N^{-1}\sum_k e^{\ii\vp_k(t)}$.
In the thermodynamic limit, the field $Z$ is periodic $Z=R\exp[\ii\bar{\w} t+\theta_0]$, where $R$ is obtained according to \eqref{eq:kop}. Thus, $U(t)=R\cos(\bar{\w}t+\theta_0)$. The ACF function of this periodic process is, according to
Eq.~\eqref{eq:acf}, $\Gamma(t)=(R^2/2)\cos\bar\w t$. Thus, the WOP calculated according to Eq.~\eqref{eq:wf}, is $W=R^4/8$. 
This value should be compared with the variance $\text{Var}(U)=\Gamma(0)=R^2/2$ of the process $U(t)$, which is 
used in the traditional characterization of the synchronization transition. Therefore, in the plots below, we use a function of the variance $V=[\text{Var}(U)]^2/2$ because, in the thermodynamic limit, it coincides with WOP $W$.

\begin{figure}[!htb]
\centering
\includegraphics[width=0.9\columnwidth]{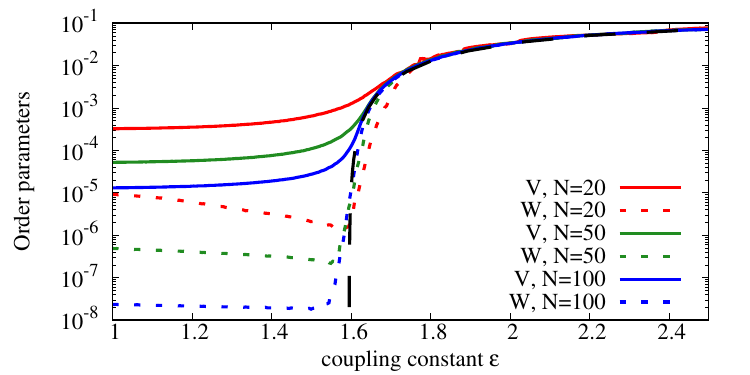}
\caption{The variance-based order parameter $V$ (solid curves) and the WOP $W$ (short-dashed curves with the corresponding colors), both for the complete observable $U(t)$, for the Kuramoto model \eqref{eq:kur} with $N=20,50,100$. The long-dashed black line shows the theoretical prediction in the thermodynamic limit \eqref{eq:kop}. One can see that the contrast between the synchronous and asynchronous states increases, if WOP is used, by several orders for $N=100$, and by more than one order for $N=20$.
Parameters of the WOP calculation: $T=10^5$, $\Tau_1=250$, $\Tau_2=500$.}
\label{fig:kur-1}
\end{figure}

We start with full global observables $U(t)$ of a population of Kuramoto oscillators and report the order parameters close to the synchronization transition in Fig.~\ref{fig:kur-1}, for system sizes $N=20,\,50,\,100$. One can see that the ``sharpness'' of the transition to synchrony is significantly improved if the WOP is used instead of the variance-based characterization. Remarkably, the WOP decreases prior to the transition threshold for a small number of units $N=20$ (this effect is also present for $N=50$ but not so pronounced). We attribute this to the chaoticity of the Kuramoto model, studied in Refs.~\onlinecite{popovych2005phase,maistrenko2005chaotic,carlu2018origin}. In these papers, it has been shown that at finite coupling strengths prior to the synchronization transition, the largest Lyapunov exponent in the Kuramoto model is positive $\sim N^{-1}$ and reaches a maximal value close to the transition. Because the chaoticity of the population makes the observed mean field $U(t)$ irregular, its ACF decreases faster for stronger chaos, and the WOP is smaller. This explains why, for small $N$, the WOP has a minimum prior to the transition to synchrony. 

\begin{figure}[!htb]
\centering
\includegraphics[width=\columnwidth]{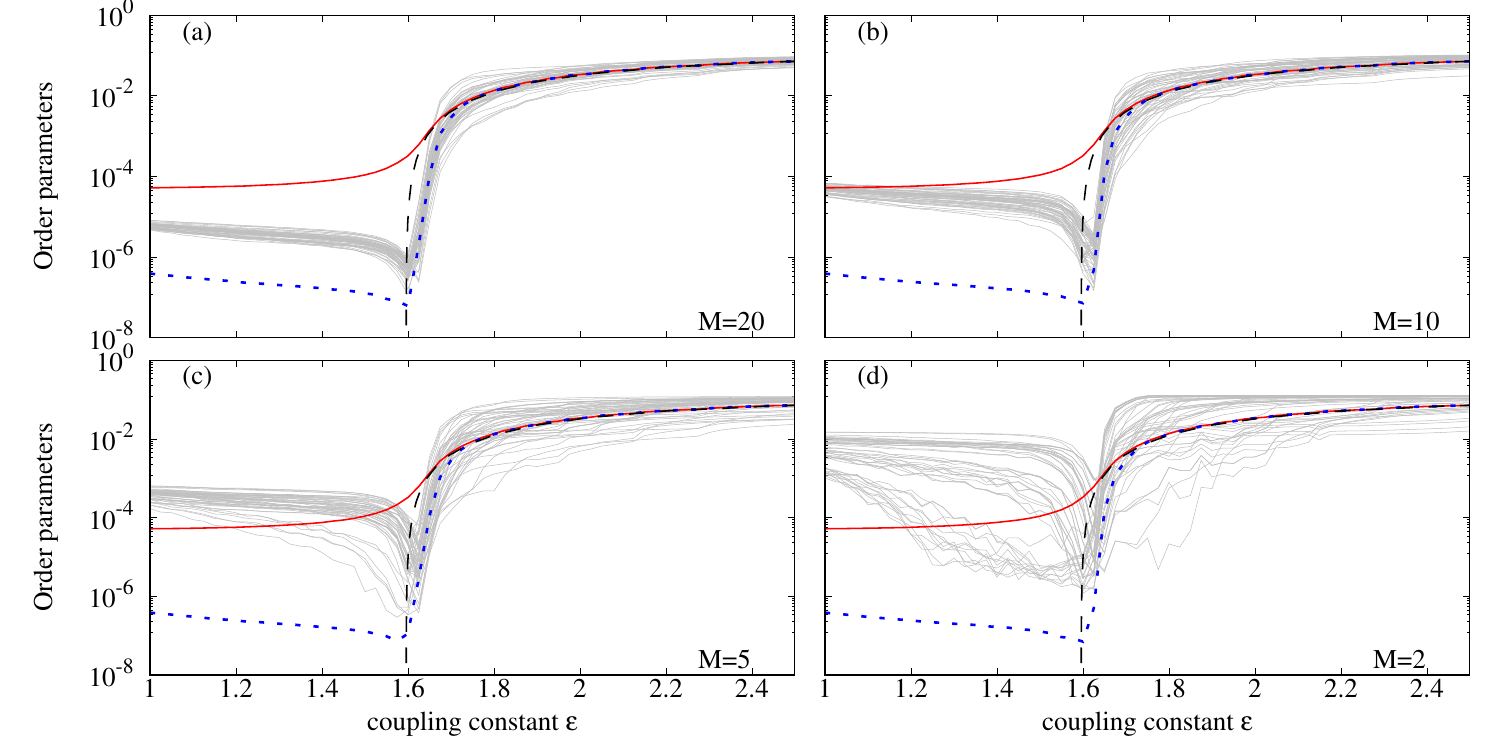}
\caption{Analysis of the synchronization transition for $N=50$ Kuramoto oscillators via observations of $M=20$ (a), $M=10$ (b), $M=5$ (c), and $M=2$ (d) units. For each case, together with the variance-based parameter $V$ (based on the full observation; solid red lines) and the WOP $W$ (dashed blue lines) calculated from the full mean field, we show WOPs for 50 random samples of $M$ selected for observation units with grey lines. The black long-dashed line is the theoretical prediction for the thermodynamic limit.
Parameters of the WOP calculation: $T=10^5$, $\Tau_=50$, $\Tau_2=100$.}
\label{fig:kur-2}
\end{figure}

Next, we explore in Fig.~\ref{fig:kur-2} how the synchronization transition can be characterized with finite samples from the ensemble of $N=50$ units. Because the oscillators are not identical (they have different natural frequencies), their behavior at the transition is different. The oscillators with the natural frequencies close to the middle one become locked by the mean field, while those with frequencies at the tails of the distribution remain nearly quasiperiodic (although they possess a periodic component from the mean field driving). Thus, we considered not just one sample of $M$ units to calculate the observed mean field $U_M(t)$ but performed calculations with a set of 50 randomly chosen samples. The results for all these 50 samples are shown in Fig.~\ref{fig:kur-2} with grey lines to illustrate the diversity of the transition characterizations due to the units' non-identity.  
One can see that some, but not all, cases with $M=2$ (panel (d)) show a reasonable contrast between asynchrony and synchrony. On the contrary, already an observation of $M=5$ units (panel (c)) 
provides a reliable description of the transition. The quality grows with $M$, and at $M=10$ (panel (b)), the quality of WOP is almost the same as the quality of the variance-based characterization where all units are used. For $M=20$ (panel (a)), the results from partial samplings are close to those with the full observability $M=50$.

One remark is in order. Because the Kuramoto model \eqref{eq:kur} is invariant with respect to time-dependent phase shifts $\vp\to\vp+\nu t$, one often performs such a transformation with $\nu=\bar{\w}$. As a result, in the transformed phases, the mean fields are not periodic but time-independent. In terms of the original oscillators, Eqs.~(\ref{eq:geq-1}-\ref{eq:geq-3}), this corresponds to mean fields in definitions of which (see Eq.~\eqref{eq:geq-3}) an explicit periodic time dependence is present. Thus, the example of the Kuramoto system does not contradict the general considerations of Section~\ref{sec:gc}, where observables that are not explicitly time-dependent are assumed.

The conclusion from this example is that although the individual oscillators are regular,
the mean fields and the fields from the samples with sufficiently large size $M$ fluctuate in the asynchronous state and have a strong periodic component in the synchronous regime. This property allows for a nice separation of synchrony and asynchrony by virtue of the WOP. However, in contradistinction to the case of local irregular units, the efficacy of the WOP approach drops at small sample sizes $M$.

\section{Conclusion}
\label{sec:con}

To summarize, we suggested an approach of synchrony quantification based on the characterizing regularity of the mean fields in a population of globally coupled units. The main difference to previous works is that we do not simply identify the existence of macroscopic mean field oscillations by following the variance but select the regular part of these oscillations by virtue of Wiener's formula applied to the auto-correlation function. We coin the term ``Wiener order parameter'' for the corresponding quantity. As we show with different examples, this approach strongly increases the contrast between the synchronous and asynchronous states. Most successful is the application of this approach to ensembles of highly irregular units, like excitable systems exhibiting noise-induced oscillation  (Section~\ref{sec:ar}) and chaotic oscillators (Section~\ref{sec:ca}). Furthermore, because in our method, we effectively separate the regularity and irregularity of the observations by means of the ACF calculations, there is, in fact, no need for ensemble averaging. This allows for working with small samples, and in most cases, even observing one unit out of the large ensemble allows for a reliable quantification of the synchronization transition.

In our approach, we operate with a generic local observable that represents the system's state. We stress that this observable does not need to be the phase of the oscillations, and there is no need to extract the phase from this observable. Of course, in the cases where the phase-dynamics description is adequate, knowing the phases would lead to an enhanced characterization of the synchronization transition by including properties of frequency entrainment, etc. The basic requirement for the observable is that it is suitable for calculations of the ACF. This means that sparse in time observables, like spikes resulting in a point process, require a special consideration to be reported elsewhere. 

While the description of synchrony/asynchrony contrast as that of regularity/irregularity is valid in many circumstances, there are situations where synchronous states show a large degree of irregularity (even in the thermodynamic limit). Irregularity of the mean fields can also result from a common noisy forcing. In such situations, the method should be significantly modified; this is a work in progress.

\acknowledgments
We thank O. Omelchenko and K. Krischer for fruitful discussions. 

\section*{Data availability}
All numerical experiments  are described in the paper and can be reproduced without additional information.


\begin{thebibliography}{46}%
\makeatletter
\providecommand \@ifxundefined [1]{%
 \@ifx{#1\undefined}
}%
\providecommand \@ifnum [1]{%
 \ifnum #1\expandafter \@firstoftwo
 \else \expandafter \@secondoftwo
 \fi
}%
\providecommand \@ifx [1]{%
 \ifx #1\expandafter \@firstoftwo
 \else \expandafter \@secondoftwo
 \fi
}%
\providecommand \natexlab [1]{#1}%
\providecommand \enquote  [1]{``#1''}%
\providecommand \bibnamefont  [1]{#1}%
\providecommand \bibfnamefont [1]{#1}%
\providecommand \citenamefont [1]{#1}%
\providecommand \href@noop [0]{\@secondoftwo}%
\providecommand \href [0]{\begingroup \@sanitize@url \@href}%
\providecommand \@href[1]{\@@startlink{#1}\@@href}%
\providecommand \@@href[1]{\endgroup#1\@@endlink}%
\providecommand \@sanitize@url [0]{\catcode `\\12\catcode `\$12\catcode
  `\&12\catcode `\#12\catcode `\^12\catcode `\_12\catcode `\%12\relax}%
\providecommand \@@startlink[1]{}%
\providecommand \@@endlink[0]{}%
\providecommand \url  [0]{\begingroup\@sanitize@url \@url }%
\providecommand \@url [1]{\endgroup\@href {#1}{\urlprefix }}%
\providecommand \urlprefix  [0]{URL }%
\providecommand \Eprint [0]{\href }%
\providecommand \doibase [0]{http://dx.doi.org/}%
\providecommand \selectlanguage [0]{\@gobble}%
\providecommand \bibinfo  [0]{\@secondoftwo}%
\providecommand \bibfield  [0]{\@secondoftwo}%
\providecommand \translation [1]{[#1]}%
\providecommand \BibitemOpen [0]{}%
\providecommand \bibitemStop [0]{}%
\providecommand \bibitemNoStop [0]{.\EOS\space}%
\providecommand \EOS [0]{\spacefactor3000\relax}%
\providecommand \BibitemShut  [1]{\csname bibitem#1\endcsname}%
\let\auto@bib@innerbib\@empty
\bibitem [{\citenamefont {Winfree}(1967)}]{Winfree-67}%
  \BibitemOpen
  \bibfield  {author} {\bibinfo {author} {\bibfnamefont {A.~T.}\ \bibnamefont
  {Winfree}},\ }\bibfield  {title} {\enquote {\bibinfo {title} {Biological
  rhythms and the behavior of populations of coupled oscillators},}\
  }\href@noop {} {\bibfield  {journal} {\bibinfo  {journal} {J. Theor. Biol.}\
  }\textbf {\bibinfo {volume} {16}},\ \bibinfo {pages} {15--42} (\bibinfo
  {year} {1967})}\BibitemShut {NoStop}%
\bibitem [{\citenamefont {Kuramoto}(1975)}]{Kuramoto-75}%
  \BibitemOpen
  \bibfield  {author} {\bibinfo {author} {\bibfnamefont {Y.}~\bibnamefont
  {Kuramoto}},\ }\bibfield  {title} {\enquote {\bibinfo {title}
  {Self-entrainment of a population of coupled nonlinear oscillators},}\ }in\
  \href@noop {} {\emph {\bibinfo {booktitle} {International Symposium on
  Mathematical Problems in Theoretical Physics}}},\ \bibinfo {editor} {edited
  by\ \bibinfo {editor} {\bibfnamefont {H.}~\bibnamefont {Araki}}}\ (\bibinfo
  {publisher} {Springer Lecture Notes Phys., v. 39},\ \bibinfo {address} {New
  York},\ \bibinfo {year} {1975})\ p.\ \bibinfo {pages} {420}\BibitemShut
  {NoStop}%
\bibitem [{\citenamefont {Strogatz}\ and\ \citenamefont
  {Stewart}(1993)}]{Strogatz-Stewart-93}%
  \BibitemOpen
  \bibfield  {author} {\bibinfo {author} {\bibfnamefont {S.~H.}\ \bibnamefont
  {Strogatz}}\ and\ \bibinfo {author} {\bibfnamefont {I.}~\bibnamefont
  {Stewart}},\ }\bibfield  {title} {\enquote {\bibinfo {title} {Coupled
  oscillators and biological synchronization},}\ }\href@noop {} {\bibfield
  {journal} {\bibinfo  {journal} {Scientific American}\ ,\ \bibinfo {pages}
  {68--75}} (\bibinfo {year} {1993})}\BibitemShut {NoStop}%
\bibitem [{\citenamefont {Acebr{\'o}n}\ \emph {et~al.}(1998)\citenamefont
  {Acebr{\'o}n}, \citenamefont {Bonilla}, \citenamefont {Leo},\ and\
  \citenamefont {Spigler}}]{Acebron-Bonilla-DeLeo-Spigler-98}%
  \BibitemOpen
  \bibfield  {author} {\bibinfo {author} {\bibfnamefont {J.~A.}\ \bibnamefont
  {Acebr{\'o}n}}, \bibinfo {author} {\bibfnamefont {L.~L.}\ \bibnamefont
  {Bonilla}}, \bibinfo {author} {\bibfnamefont {S.~D.}\ \bibnamefont {Leo}}, \
  and\ \bibinfo {author} {\bibfnamefont {R.}~\bibnamefont {Spigler}},\
  }\bibfield  {title} {\enquote {\bibinfo {title} {Breaking the symmetry in
  bimodal frequency distributions of globally coupled oscillators},}\
  }\href@noop {} {\bibfield  {journal} {\bibinfo  {journal} {Phys. Rev. E}\
  }\textbf {\bibinfo {volume} {57}},\ \bibinfo {pages} {5287--5290} (\bibinfo
  {year} {1998})}\BibitemShut {NoStop}%
\bibitem [{\citenamefont {Pikovsky}, \citenamefont {Rosenblum},\ and\
  \citenamefont {Kurths}(2001)}]{Pikovsky-Rosenblum-Kurths-01}%
  \BibitemOpen
  \bibfield  {author} {\bibinfo {author} {\bibfnamefont {A.}~\bibnamefont
  {Pikovsky}}, \bibinfo {author} {\bibfnamefont {M.}~\bibnamefont {Rosenblum}},
  \ and\ \bibinfo {author} {\bibfnamefont {J.}~\bibnamefont {Kurths}},\
  }\href@noop {} {\emph {\bibinfo {title} {Synchronization. A Universal Concept
  in Nonlinear Sciences.}}}\ (\bibinfo  {publisher} {Cambridge University
  Press},\ \bibinfo {address} {Cambridge},\ \bibinfo {year} {2001})\BibitemShut
  {NoStop}%
\bibitem [{\citenamefont {Eckhardt}\ \emph {et~al.}(2007)\citenamefont
  {Eckhardt}, \citenamefont {Ott}, \citenamefont {Strogatz}, \citenamefont
  {Abrams},\ and\ \citenamefont {McRobie}}]{Eckhardt_et_al-07}%
  \BibitemOpen
  \bibfield  {author} {\bibinfo {author} {\bibfnamefont {B.}~\bibnamefont
  {Eckhardt}}, \bibinfo {author} {\bibfnamefont {E.}~\bibnamefont {Ott}},
  \bibinfo {author} {\bibfnamefont {S.~H.}\ \bibnamefont {Strogatz}}, \bibinfo
  {author} {\bibfnamefont {D.~M.}\ \bibnamefont {Abrams}}, \ and\ \bibinfo
  {author} {\bibfnamefont {A.}~\bibnamefont {McRobie}},\ }\bibfield  {title}
  {\enquote {\bibinfo {title} {Modeling walker synchronization on the
  {M}illennium {B}ridge},}\ }\href@noop {} {\bibfield  {journal} {\bibinfo
  {journal} {Phys. Rev. E}\ }\textbf {\bibinfo {volume} {75}},\ \bibinfo
  {pages} {021110} (\bibinfo {year} {2007})}\BibitemShut {NoStop}%
\bibitem [{\citenamefont {Glass}\ and\ \citenamefont
  {Mackey}(1988)}]{Glass-Mackey-88}%
  \BibitemOpen
  \bibfield  {author} {\bibinfo {author} {\bibfnamefont {L.}~\bibnamefont
  {Glass}}\ and\ \bibinfo {author} {\bibfnamefont {M.~C.}\ \bibnamefont
  {Mackey}},\ }\href@noop {} {\emph {\bibinfo {title} {From Clocks to Chaos:
  {T}he Rhythms of Life.}}}\ (\bibinfo  {publisher} {Princeton Univ. Press},\
  \bibinfo {address} {Princeton, NJ},\ \bibinfo {year} {1988})\BibitemShut
  {NoStop}%
\bibitem [{\citenamefont {Buzs{\'a}ki}(2006)}]{Buzsaki-06}%
  \BibitemOpen
  \bibfield  {author} {\bibinfo {author} {\bibfnamefont {G.}~\bibnamefont
  {Buzs{\'a}ki}},\ }\href@noop {} {\emph {\bibinfo {title} {Rhythms of the
  brain}}}\ (\bibinfo  {publisher} {Oxford UP},\ \bibinfo {address} {Oxford},\
  \bibinfo {year} {2006})\BibitemShut {NoStop}%
\bibitem [{\citenamefont {Lehnertz}(2008)}]{lehnertz2008epilepsy}%
  \BibitemOpen
  \bibfield  {author} {\bibinfo {author} {\bibfnamefont {K.}~\bibnamefont
  {Lehnertz}},\ }\bibfield  {title} {\enquote {\bibinfo {title} {Epilepsy and
  nonlinear dynamics},}\ }\href@noop {} {\bibfield  {journal} {\bibinfo
  {journal} {Journal of Biological Physics}\ }\textbf {\bibinfo {volume}
  {34}},\ \bibinfo {pages} {253--266} (\bibinfo {year} {2008})}\BibitemShut
  {NoStop}%
\bibitem [{\citenamefont {Daffertshofer}\ and\ \citenamefont
  {Pietras}(2020)}]{daffertshofer2020phase}%
  \BibitemOpen
  \bibfield  {author} {\bibinfo {author} {\bibfnamefont {A.}~\bibnamefont
  {Daffertshofer}}\ and\ \bibinfo {author} {\bibfnamefont {B.}~\bibnamefont
  {Pietras}},\ }\bibfield  {title} {\enquote {\bibinfo {title} {Phase
  synchronization in neural systems},}\ }\href@noop {} {\bibfield  {journal}
  {\bibinfo  {journal} {Synergetics}\ ,\ \bibinfo {pages} {221--233}} (\bibinfo
  {year} {2020})}\BibitemShut {NoStop}%
\bibitem [{\citenamefont {Ginzburg}\ and\ \citenamefont
  {Sompolinsky}(1994)}]{ginzburg1994theory}%
  \BibitemOpen
  \bibfield  {author} {\bibinfo {author} {\bibfnamefont {I.}~\bibnamefont
  {Ginzburg}}\ and\ \bibinfo {author} {\bibfnamefont {H.}~\bibnamefont
  {Sompolinsky}},\ }\bibfield  {title} {\enquote {\bibinfo {title} {Theory of
  correlations in stochastic neural networks},}\ }\href@noop {} {\bibfield
  {journal} {\bibinfo  {journal} {Physical Review E}\ }\textbf {\bibinfo
  {volume} {50}},\ \bibinfo {pages} {3171} (\bibinfo {year}
  {1994})}\BibitemShut {NoStop}%
\bibitem [{\citenamefont {Hansel}\ and\ \citenamefont
  {Sompolinsky}(1996)}]{hansel1996chaos}%
  \BibitemOpen
  \bibfield  {author} {\bibinfo {author} {\bibfnamefont {D.}~\bibnamefont
  {Hansel}}\ and\ \bibinfo {author} {\bibfnamefont {H.}~\bibnamefont
  {Sompolinsky}},\ }\bibfield  {title} {\enquote {\bibinfo {title} {Chaos and
  synchrony in a model of a hypercolumn in visual cortex},}\ }\href@noop {}
  {\bibfield  {journal} {\bibinfo  {journal} {Journal of Computational
  Neuroscience}\ }\textbf {\bibinfo {volume} {3}},\ \bibinfo {pages} {7--34}
  (\bibinfo {year} {1996})}\BibitemShut {NoStop}%
\bibitem [{\citenamefont {Pikovsky}, \citenamefont {Rosenblum},\ and\
  \citenamefont {Kurths}(1996)}]{Pikovsky-Rosenblum-Kurths-96}%
  \BibitemOpen
  \bibfield  {author} {\bibinfo {author} {\bibfnamefont {A.}~\bibnamefont
  {Pikovsky}}, \bibinfo {author} {\bibfnamefont {M.}~\bibnamefont {Rosenblum}},
  \ and\ \bibinfo {author} {\bibfnamefont {J.}~\bibnamefont {Kurths}},\
  }\bibfield  {title} {\enquote {\bibinfo {title} {Synchronization in a
  population of globally coupled chaotic oscillators},}\ }\href@noop {}
  {\bibfield  {journal} {\bibinfo  {journal} {Europhys. Lett.}\ }\textbf
  {\bibinfo {volume} {34}},\ \bibinfo {pages} {165--170} (\bibinfo {year}
  {1996})}\BibitemShut {NoStop}%
\bibitem [{\citenamefont {Golomb}, \citenamefont {Hansel},\ and\ \citenamefont
  {Mato}(2001)}]{golomb2001mechanisms}%
  \BibitemOpen
  \bibfield  {author} {\bibinfo {author} {\bibfnamefont {D.}~\bibnamefont
  {Golomb}}, \bibinfo {author} {\bibfnamefont {D.}~\bibnamefont {Hansel}}, \
  and\ \bibinfo {author} {\bibfnamefont {G.}~\bibnamefont {Mato}},\ }\bibfield
  {title} {\enquote {\bibinfo {title} {Mechanisms of synchrony of neural
  activity in large networks},}\ }in\ \href@noop {} {\emph {\bibinfo
  {booktitle} {Handbook of Biological Physics}}},\ Vol.~\bibinfo {volume} {4}\
  (\bibinfo  {publisher} {Elsevier},\ \bibinfo {year} {2001})\ pp.\ \bibinfo
  {pages} {887--968}\BibitemShut {NoStop}%
\bibitem [{\citenamefont {Kapitaniak}\ \emph {et~al.}(2012)\citenamefont
  {Kapitaniak}, \citenamefont {Czolczy{\'n}ski}, \citenamefont {Perlikowski},
  \citenamefont {Stefa{\'n}ski},\ and\ \citenamefont
  {Kapitaniak}}]{Kapitaniak_etal-12}%
  \BibitemOpen
  \bibfield  {author} {\bibinfo {author} {\bibfnamefont {M.}~\bibnamefont
  {Kapitaniak}}, \bibinfo {author} {\bibfnamefont {K.}~\bibnamefont
  {Czolczy{\'n}ski}}, \bibinfo {author} {\bibfnamefont {P.}~\bibnamefont
  {Perlikowski}}, \bibinfo {author} {\bibfnamefont {A.}~\bibnamefont
  {Stefa{\'n}ski}}, \ and\ \bibinfo {author} {\bibfnamefont {T.}~\bibnamefont
  {Kapitaniak}},\ }\bibfield  {title} {\enquote {\bibinfo {title}
  {Synchronization of clocks},}\ }\href@noop {} {\bibfield  {journal} {\bibinfo
   {journal} {Physics Reports}\ }\textbf {\bibinfo {volume} {517}},\ \bibinfo
  {pages} {1--69} (\bibinfo {year} {2012})}\BibitemShut {NoStop}%
\bibitem [{\citenamefont {Czolczy{\'n}ski}\ \emph {et~al.}(2013)\citenamefont
  {Czolczy{\'n}ski}, \citenamefont {Perlikowski}, \citenamefont
  {Stefa{\'n}ski},\ and\ \citenamefont {Kapitaniak}}]{Czolczynski_etal-13}%
  \BibitemOpen
  \bibfield  {author} {\bibinfo {author} {\bibfnamefont {K.}~\bibnamefont
  {Czolczy{\'n}ski}}, \bibinfo {author} {\bibfnamefont {P.}~\bibnamefont
  {Perlikowski}}, \bibinfo {author} {\bibfnamefont {A.}~\bibnamefont
  {Stefa{\'n}ski}}, \ and\ \bibinfo {author} {\bibfnamefont {T.}~\bibnamefont
  {Kapitaniak}},\ }\bibfield  {title} {\enquote {\bibinfo {title}
  {Synchronization of the self-excited pendula suspended on the vertically
  displacing beam},}\ }\href@noop {} {\bibfield  {journal} {\bibinfo  {journal}
  {Communications in Nonlinear Science and Numerical Simulation}\ }\textbf
  {\bibinfo {volume} {18}},\ \bibinfo {pages} {386 -- 400} (\bibinfo {year}
  {2013})}\BibitemShut {NoStop}%
\bibitem [{\citenamefont {Wiesenfeld}, \citenamefont {Colet},\ and\
  \citenamefont {Strogatz}(1998)}]{Wiesenfeld-Colet-Strogatz-98}%
  \BibitemOpen
  \bibfield  {author} {\bibinfo {author} {\bibfnamefont {K.}~\bibnamefont
  {Wiesenfeld}}, \bibinfo {author} {\bibfnamefont {P.}~\bibnamefont {Colet}}, \
  and\ \bibinfo {author} {\bibfnamefont {S.}~\bibnamefont {Strogatz}},\
  }\bibfield  {title} {\enquote {\bibinfo {title} {{Frequency locking in
  Josephson arrays: Connection with the Kuramoto model}},}\ }\href@noop {}
  {\bibfield  {journal} {\bibinfo  {journal} {Physical Review E}\ }\textbf
  {\bibinfo {volume} {57}},\ \bibinfo {pages} {1563--1569} (\bibinfo {year}
  {1998})}\BibitemShut {NoStop}%
\bibitem [{\citenamefont {Temirbayev}\ \emph {et~al.}(2012)\citenamefont
  {Temirbayev}, \citenamefont {Zhanabaev}, \citenamefont {Tarasov},
  \citenamefont {Ponomarenko},\ and\ \citenamefont
  {Rosenblum}}]{Temirbayev_etal-12}%
  \BibitemOpen
  \bibfield  {author} {\bibinfo {author} {\bibfnamefont {A.~A.}\ \bibnamefont
  {Temirbayev}}, \bibinfo {author} {\bibfnamefont {Z.~Z.}\ \bibnamefont
  {Zhanabaev}}, \bibinfo {author} {\bibfnamefont {S.~B.}\ \bibnamefont
  {Tarasov}}, \bibinfo {author} {\bibfnamefont {V.~I.}\ \bibnamefont
  {Ponomarenko}}, \ and\ \bibinfo {author} {\bibfnamefont {M.}~\bibnamefont
  {Rosenblum}},\ }\bibfield  {title} {\enquote {\bibinfo {title} {Experiments
  on oscillator ensembles with global nonlinear coupling},}\ }\href@noop {}
  {\bibfield  {journal} {\bibinfo  {journal} {Phys. Rev. E}\ }\textbf {\bibinfo
  {volume} {85}},\ \bibinfo {pages} {015204} (\bibinfo {year}
  {2012})}\BibitemShut {NoStop}%
\bibitem [{\citenamefont {Wenzel}\ and\ \citenamefont
  {Hamm}(2021)}]{wenzel2021identification}%
  \BibitemOpen
  \bibfield  {author} {\bibinfo {author} {\bibfnamefont {M.}~\bibnamefont
  {Wenzel}}\ and\ \bibinfo {author} {\bibfnamefont {J.~P.}\ \bibnamefont
  {Hamm}},\ }\bibfield  {title} {\enquote {\bibinfo {title} {Identification and
  quantification of neuronal ensembles in optical imaging experiments},}\
  }\href@noop {} {\bibfield  {journal} {\bibinfo  {journal} {Journal of
  Neuroscience Methods}\ }\textbf {\bibinfo {volume} {351}},\ \bibinfo {pages}
  {109046} (\bibinfo {year} {2021})}\BibitemShut {NoStop}%
\bibitem [{Note1()}]{Note1}%
  \BibitemOpen
  \bibinfo {note} {In this paper we assume ergodicity and stationarity, so
  practically the averaging is performed over time.}\BibitemShut {Stop}%
\bibitem [{Note2()}]{Note2}%
  \BibitemOpen
  \bibinfo {note} {In the calculation of the ACF it is usual to assume that the
  regular component possesses a random phase shift}\BibitemShut {NoStop}%
\bibitem [{\citenamefont {Wiener}(1930)}]{wiener1930generalized}%
  \BibitemOpen
  \bibfield  {author} {\bibinfo {author} {\bibfnamefont {N.}~\bibnamefont
  {Wiener}},\ }\bibfield  {title} {\enquote {\bibinfo {title} {Generalized
  harmonic analysis},}\ }\href@noop {} {\bibfield  {journal} {\bibinfo
  {journal} {Acta mathematica}\ }\textbf {\bibinfo {volume} {55}},\ \bibinfo
  {pages} {117--258} (\bibinfo {year} {1930})}\BibitemShut {NoStop}%
\bibitem [{\citenamefont {Kornfeld}, \citenamefont {Sinai},\ and\ \citenamefont
  {Fomin}(1982)}]{kornfeld1982ergodic}%
  \BibitemOpen
  \bibfield  {author} {\bibinfo {author} {\bibfnamefont {I.}~\bibnamefont
  {Kornfeld}}, \bibinfo {author} {\bibfnamefont {Y.~G.}\ \bibnamefont {Sinai}},
  \ and\ \bibinfo {author} {\bibfnamefont {S.}~\bibnamefont {Fomin}},\
  }\href@noop {} {\emph {\bibinfo {title} {Ergodic theory, Nauka, Moscow, 1980;
  English transl., Springer, New York}}}\ (\bibinfo {year} {1982})\BibitemShut
  {NoStop}%
\bibitem [{\citenamefont {Queff{\'e}lec}(2010)}]{queffelec2010substitution}%
  \BibitemOpen
  \bibfield  {author} {\bibinfo {author} {\bibfnamefont {M.}~\bibnamefont
  {Queff{\'e}lec}},\ }\href@noop {} {\emph {\bibinfo {title} {Substitution
  dynamical systems-spectral analysis}}},\ \bibinfo {series} {Lecture Notes in
  Mathematics}, Vol.\ \bibinfo {volume} {1294}\ (\bibinfo  {publisher}
  {Springer},\ \bibinfo {year} {2010})\BibitemShut {NoStop}%
\bibitem [{\citenamefont {Daido}(1990)}]{Daido-90}%
  \BibitemOpen
  \bibfield  {author} {\bibinfo {author} {\bibfnamefont {H.}~\bibnamefont
  {Daido}},\ }\bibfield  {title} {\enquote {\bibinfo {title} {Intrinsic
  fluctuations and a phase transition in a class of large population of
  interacting oscillators},}\ }\href@noop {} {\bibfield  {journal} {\bibinfo
  {journal} {J. Stat. Phys.}\ }\textbf {\bibinfo {volume} {60}},\ \bibinfo
  {pages} {753--800} (\bibinfo {year} {1990})}\BibitemShut {NoStop}%
\bibitem [{\citenamefont {Pikovsky}\ and\ \citenamefont
  {Ruffo}(1999)}]{Pikovsky-Ruffo-99}%
  \BibitemOpen
  \bibfield  {author} {\bibinfo {author} {\bibfnamefont {A.}~\bibnamefont
  {Pikovsky}}\ and\ \bibinfo {author} {\bibfnamefont {S.}~\bibnamefont
  {Ruffo}},\ }\bibfield  {title} {\enquote {\bibinfo {title} {Finite-size
  effects in a population of interacting oscillators},}\ }\href@noop {}
  {\bibfield  {journal} {\bibinfo  {journal} {Phys. Rev. E}\ }\textbf {\bibinfo
  {volume} {59}},\ \bibinfo {pages} {1633--1636} (\bibinfo {year}
  {1999})}\BibitemShut {NoStop}%
\bibitem [{\citenamefont {Bertini}, \citenamefont {Giacomin},\ and\
  \citenamefont {Poquet}(2014)}]{bertini2014synchronization}%
  \BibitemOpen
  \bibfield  {author} {\bibinfo {author} {\bibfnamefont {L.}~\bibnamefont
  {Bertini}}, \bibinfo {author} {\bibfnamefont {G.}~\bibnamefont {Giacomin}}, \
  and\ \bibinfo {author} {\bibfnamefont {C.}~\bibnamefont {Poquet}},\
  }\bibfield  {title} {\enquote {\bibinfo {title} {Synchronization and random
  long time dynamics for mean-field plane rotators},}\ }\href@noop {}
  {\bibfield  {journal} {\bibinfo  {journal} {Probability Theory and Related
  Fields}\ }\textbf {\bibinfo {volume} {160}},\ \bibinfo {pages} {593--653}
  (\bibinfo {year} {2014})}\BibitemShut {NoStop}%
\bibitem [{\citenamefont {Hong}\ \emph {et~al.}(2015)\citenamefont {Hong},
  \citenamefont {Chat\'e}, \citenamefont {Tang},\ and\ \citenamefont
  {Park}}]{Hong-15}%
  \BibitemOpen
  \bibfield  {author} {\bibinfo {author} {\bibfnamefont {H.}~\bibnamefont
  {Hong}}, \bibinfo {author} {\bibfnamefont {H.}~\bibnamefont {Chat\'e}},
  \bibinfo {author} {\bibfnamefont {L.-H.}\ \bibnamefont {Tang}}, \ and\
  \bibinfo {author} {\bibfnamefont {H.}~\bibnamefont {Park}},\ }\bibfield
  {title} {\enquote {\bibinfo {title} {Finite-size scaling, dynamic
  fluctuations, and hyperscaling relation in the {K}uramoto model},}\ }\href
  {\doibase 10.1103/PhysRevE.92.022122} {\bibfield  {journal} {\bibinfo
  {journal} {Phys. Rev. E}\ }\textbf {\bibinfo {volume} {92}},\ \bibinfo
  {pages} {022122} (\bibinfo {year} {2015})}\BibitemShut {NoStop}%
\bibitem [{\citenamefont {Gottwald}(2017)}]{gottwald2017finite}%
  \BibitemOpen
  \bibfield  {author} {\bibinfo {author} {\bibfnamefont {G.~A.}\ \bibnamefont
  {Gottwald}},\ }\bibfield  {title} {\enquote {\bibinfo {title} {Finite-size
  effects in a stochastic {K}uramoto model},}\ }\href@noop {} {\bibfield
  {journal} {\bibinfo  {journal} {Chaos: An Interdisciplinary Journal of
  Nonlinear Science}\ }\textbf {\bibinfo {volume} {27}} (\bibinfo {year}
  {2017})}\BibitemShut {NoStop}%
\bibitem [{\citenamefont {Peter}\ and\ \citenamefont
  {Pikovsky}(2018)}]{Peter-Pikovsky-18}%
  \BibitemOpen
  \bibfield  {author} {\bibinfo {author} {\bibfnamefont {F.}~\bibnamefont
  {Peter}}\ and\ \bibinfo {author} {\bibfnamefont {A.}~\bibnamefont
  {Pikovsky}},\ }\bibfield  {title} {\enquote {\bibinfo {title} {Transition to
  collective oscillations in finite {K}uramoto ensembles},}\ }\href@noop {}
  {\bibfield  {journal} {\bibinfo  {journal} {Phys. Rev. E}\ }\textbf {\bibinfo
  {volume} {97}},\ \bibinfo {pages} {032310} (\bibinfo {year}
  {2018})}\BibitemShut {NoStop}%
\bibitem [{\citenamefont {Yue}\ and\ \citenamefont
  {Gottwald}(2023)}]{yue2023stochastic}%
  \BibitemOpen
  \bibfield  {author} {\bibinfo {author} {\bibfnamefont {W.}~\bibnamefont
  {Yue}}\ and\ \bibinfo {author} {\bibfnamefont {G.~A.}\ \bibnamefont
  {Gottwald}},\ }\href@noop {} {\enquote {\bibinfo {title} {A stochastic
  approximation for the finite-size {K}uramoto-{S}akaguchi model},}\ }
  (\bibinfo {year} {2023}),\ \Eprint {http://arxiv.org/abs/2310.20048}
  {arXiv:2310.20048 [nlin.AO]} \BibitemShut {NoStop}%
\bibitem [{\citenamefont {Hakim}\ and\ \citenamefont
  {Rappel}(1992)}]{hakim1992dynamics}%
  \BibitemOpen
  \bibfield  {author} {\bibinfo {author} {\bibfnamefont {V.}~\bibnamefont
  {Hakim}}\ and\ \bibinfo {author} {\bibfnamefont {W.-J.}\ \bibnamefont
  {Rappel}},\ }\bibfield  {title} {\enquote {\bibinfo {title} {Dynamics of the
  globally coupled complex {G}inzburg-{L}andau equation},}\ }\href@noop {}
  {\bibfield  {journal} {\bibinfo  {journal} {Physical Review A}\ }\textbf
  {\bibinfo {volume} {46}},\ \bibinfo {pages} {R7347} (\bibinfo {year}
  {1992})}\BibitemShut {NoStop}%
\bibitem [{\citenamefont {Nakagawa}\ and\ \citenamefont
  {Kuramoto}(1994)}]{nakagawa1994collective}%
  \BibitemOpen
  \bibfield  {author} {\bibinfo {author} {\bibfnamefont {N.}~\bibnamefont
  {Nakagawa}}\ and\ \bibinfo {author} {\bibfnamefont {Y.}~\bibnamefont
  {Kuramoto}},\ }\bibfield  {title} {\enquote {\bibinfo {title} {From
  collective oscillations to collective chaos in a globally coupled oscillator
  system},}\ }\href@noop {} {\bibfield  {journal} {\bibinfo  {journal} {Physica
  D: Nonlinear Phenomena}\ }\textbf {\bibinfo {volume} {75}},\ \bibinfo {pages}
  {74--80} (\bibinfo {year} {1994})}\BibitemShut {NoStop}%
\bibitem [{\citenamefont {Chabanol}, \citenamefont {Hakim},\ and\ \citenamefont
  {Rappel}(1997)}]{chabanol1997collective}%
  \BibitemOpen
  \bibfield  {author} {\bibinfo {author} {\bibfnamefont {M.-L.}\ \bibnamefont
  {Chabanol}}, \bibinfo {author} {\bibfnamefont {V.}~\bibnamefont {Hakim}}, \
  and\ \bibinfo {author} {\bibfnamefont {W.-J.}\ \bibnamefont {Rappel}},\
  }\bibfield  {title} {\enquote {\bibinfo {title} {Collective chaos and noise
  in the globally coupled complex {G}inzburg-{L}andau equation},}\ }\href@noop
  {} {\bibfield  {journal} {\bibinfo  {journal} {Physica D: Nonlinear
  Phenomena}\ }\textbf {\bibinfo {volume} {103}},\ \bibinfo {pages} {273--293}
  (\bibinfo {year} {1997})}\BibitemShut {NoStop}%
\bibitem [{\citenamefont {Ku}, \citenamefont {Girvan},\ and\ \citenamefont
  {Ott}(2015)}]{ku2015dynamical}%
  \BibitemOpen
  \bibfield  {author} {\bibinfo {author} {\bibfnamefont {W.~L.}\ \bibnamefont
  {Ku}}, \bibinfo {author} {\bibfnamefont {M.}~\bibnamefont {Girvan}}, \ and\
  \bibinfo {author} {\bibfnamefont {E.}~\bibnamefont {Ott}},\ }\bibfield
  {title} {\enquote {\bibinfo {title} {Dynamical transitions in large systems
  of mean field-coupled {L}andau-{S}tuart oscillators: Extensive chaos and
  cluster states},}\ }\href@noop {} {\bibfield  {journal} {\bibinfo  {journal}
  {Chaos: An Interdisciplinary Journal of Nonlinear Science}\ }\textbf
  {\bibinfo {volume} {25}} (\bibinfo {year} {2015})}\BibitemShut {NoStop}%
\bibitem [{\citenamefont {Clusella}\ and\ \citenamefont
  {Politi}(2019)}]{clusella2019between}%
  \BibitemOpen
  \bibfield  {author} {\bibinfo {author} {\bibfnamefont {P.}~\bibnamefont
  {Clusella}}\ and\ \bibinfo {author} {\bibfnamefont {A.}~\bibnamefont
  {Politi}},\ }\bibfield  {title} {\enquote {\bibinfo {title} {Between phase
  and amplitude oscillators},}\ }\href@noop {} {\bibfield  {journal} {\bibinfo
  {journal} {Physical Review E}\ }\textbf {\bibinfo {volume} {99}},\ \bibinfo
  {pages} {062201} (\bibinfo {year} {2019})}\BibitemShut {NoStop}%
\bibitem [{\citenamefont {Le{\'o}n}\ and\ \citenamefont
  {Paz{\'o}}(2022)}]{leon2022enlarged}%
  \BibitemOpen
  \bibfield  {author} {\bibinfo {author} {\bibfnamefont {I.}~\bibnamefont
  {Le{\'o}n}}\ and\ \bibinfo {author} {\bibfnamefont {D.}~\bibnamefont
  {Paz{\'o}}},\ }\bibfield  {title} {\enquote {\bibinfo {title} {Enlarged
  {K}uramoto model: Secondary instability and transition to collective
  chaos},}\ }\href@noop {} {\bibfield  {journal} {\bibinfo  {journal} {Physical
  Review E}\ }\textbf {\bibinfo {volume} {105}},\ \bibinfo {pages} {L042201}
  (\bibinfo {year} {2022})}\BibitemShut {NoStop}%
\bibitem [{\citenamefont {Olmi}, \citenamefont {Politi},\ and\ \citenamefont
  {Torcini}(2011)}]{olmi2011collective}%
  \BibitemOpen
  \bibfield  {author} {\bibinfo {author} {\bibfnamefont {S.}~\bibnamefont
  {Olmi}}, \bibinfo {author} {\bibfnamefont {A.}~\bibnamefont {Politi}}, \ and\
  \bibinfo {author} {\bibfnamefont {A.}~\bibnamefont {Torcini}},\ }\bibfield
  {title} {\enquote {\bibinfo {title} {Collective chaos in pulse-coupled neural
  networks},}\ }\href@noop {} {\bibfield  {journal} {\bibinfo  {journal}
  {Europhysics Letters}\ }\textbf {\bibinfo {volume} {92}},\ \bibinfo {pages}
  {60007} (\bibinfo {year} {2011})}\BibitemShut {NoStop}%
\bibitem [{\citenamefont {Paz{\'o}}\ and\ \citenamefont
  {Montbri{\'o}}(2016)}]{pazo2016quasiperiodic}%
  \BibitemOpen
  \bibfield  {author} {\bibinfo {author} {\bibfnamefont {D.}~\bibnamefont
  {Paz{\'o}}}\ and\ \bibinfo {author} {\bibfnamefont {E.}~\bibnamefont
  {Montbri{\'o}}},\ }\bibfield  {title} {\enquote {\bibinfo {title} {From
  quasiperiodic partial synchronization to collective chaos in populations of
  inhibitory neurons with delay},}\ }\href@noop {} {\bibfield  {journal}
  {\bibinfo  {journal} {Physical Review Letters}\ }\textbf {\bibinfo {volume}
  {116}},\ \bibinfo {pages} {238101} (\bibinfo {year} {2016})}\BibitemShut
  {NoStop}%
\bibitem [{\citenamefont {Ratas}\ and\ \citenamefont
  {Pyragas}(2018)}]{ratas2018macroscopic}%
  \BibitemOpen
  \bibfield  {author} {\bibinfo {author} {\bibfnamefont {I.}~\bibnamefont
  {Ratas}}\ and\ \bibinfo {author} {\bibfnamefont {K.}~\bibnamefont
  {Pyragas}},\ }\bibfield  {title} {\enquote {\bibinfo {title} {Macroscopic
  oscillations of a quadratic integrate-and-fire neuron network with global
  distributed-delay coupling},}\ }\href@noop {} {\bibfield  {journal} {\bibinfo
   {journal} {Physical Review E}\ }\textbf {\bibinfo {volume} {98}},\ \bibinfo
  {pages} {052224} (\bibinfo {year} {2018})}\BibitemShut {NoStop}%
\bibitem [{\citenamefont {Klinshov}\ \emph {et~al.}(2021)\citenamefont
  {Klinshov}, \citenamefont {Kirillov}, \citenamefont {Nekorkin},\ and\
  \citenamefont {Wolfrum}}]{klinshov2021noise}%
  \BibitemOpen
  \bibfield  {author} {\bibinfo {author} {\bibfnamefont {V.~V.}\ \bibnamefont
  {Klinshov}}, \bibinfo {author} {\bibfnamefont {S.~Y.}\ \bibnamefont
  {Kirillov}}, \bibinfo {author} {\bibfnamefont {V.~I.}\ \bibnamefont
  {Nekorkin}}, \ and\ \bibinfo {author} {\bibfnamefont {M.}~\bibnamefont
  {Wolfrum}},\ }\bibfield  {title} {\enquote {\bibinfo {title} {Noise-induced
  dynamical regimes in a system of globally coupled excitable units},}\
  }\href@noop {} {\bibfield  {journal} {\bibinfo  {journal} {Chaos: An
  Interdisciplinary Journal of Nonlinear Science}\ }\textbf {\bibinfo {volume}
  {31}} (\bibinfo {year} {2021})}\BibitemShut {NoStop}%
\bibitem [{Note3()}]{Note3}%
  \BibitemOpen
  \bibinfo {note} {Notice that the normalization of the noise strength differs
  from that used in Eq.~(\ref {eq:ar})}\BibitemShut {NoStop}%
\bibitem [{\citenamefont {Rosenblum}\ and\ \citenamefont
  {Pikovsky}(2015)}]{Rosenblum-Pikovsky-15}%
  \BibitemOpen
  \bibfield  {author} {\bibinfo {author} {\bibfnamefont {M.}~\bibnamefont
  {Rosenblum}}\ and\ \bibinfo {author} {\bibfnamefont {A.}~\bibnamefont
  {Pikovsky}},\ }\bibfield  {title} {\enquote {\bibinfo {title} {Two types of
  quasiperiodic partial synchrony in oscillator ensembles},}\ }\href@noop {}
  {\bibfield  {journal} {\bibinfo  {journal} {Phys. Rev. E}\ }\textbf {\bibinfo
  {volume} {92}},\ \bibinfo {pages} {012919} (\bibinfo {year}
  {2015})}\BibitemShut {NoStop}%
\bibitem [{\citenamefont {Carlu}, \citenamefont {Ginelli},\ and\ \citenamefont
  {Politi}(2018)}]{carlu2018origin}%
  \BibitemOpen
  \bibfield  {author} {\bibinfo {author} {\bibfnamefont {M.}~\bibnamefont
  {Carlu}}, \bibinfo {author} {\bibfnamefont {F.}~\bibnamefont {Ginelli}}, \
  and\ \bibinfo {author} {\bibfnamefont {A.}~\bibnamefont {Politi}},\
  }\bibfield  {title} {\enquote {\bibinfo {title} {Origin and scaling of chaos
  in weakly coupled phase oscillators},}\ }\href@noop {} {\bibfield  {journal}
  {\bibinfo  {journal} {Physical Review E}\ }\textbf {\bibinfo {volume} {97}},\
  \bibinfo {pages} {012203} (\bibinfo {year} {2018})}\BibitemShut {NoStop}%
\bibitem [{\citenamefont {Popovych}, \citenamefont {Maistrenko},\ and\
  \citenamefont {Tass}(2005)}]{popovych2005phase}%
  \BibitemOpen
  \bibfield  {author} {\bibinfo {author} {\bibfnamefont {O.~V.}\ \bibnamefont
  {Popovych}}, \bibinfo {author} {\bibfnamefont {Y.~L.}\ \bibnamefont
  {Maistrenko}}, \ and\ \bibinfo {author} {\bibfnamefont {P.~A.}\ \bibnamefont
  {Tass}},\ }\bibfield  {title} {\enquote {\bibinfo {title} {Phase chaos in
  coupled oscillators},}\ }\href@noop {} {\bibfield  {journal} {\bibinfo
  {journal} {Physical Review E}\ }\textbf {\bibinfo {volume} {71}},\ \bibinfo
  {pages} {065201} (\bibinfo {year} {2005})}\BibitemShut {NoStop}%
\bibitem [{\citenamefont {Maistrenko}, \citenamefont {Popovych},\ and\
  \citenamefont {Tass}(2005)}]{maistrenko2005chaotic}%
  \BibitemOpen
  \bibfield  {author} {\bibinfo {author} {\bibfnamefont {Y.~L.}\ \bibnamefont
  {Maistrenko}}, \bibinfo {author} {\bibfnamefont {O.~V.}\ \bibnamefont
  {Popovych}}, \ and\ \bibinfo {author} {\bibfnamefont {P.~A.}\ \bibnamefont
  {Tass}},\ }\bibfield  {title} {\enquote {\bibinfo {title} {Chaotic attractor
  in the {K}uramoto model},}\ }\href@noop {} {\bibfield  {journal} {\bibinfo
  {journal} {International Journal of Bifurcation and Chaos}\ }\textbf
  {\bibinfo {volume} {15}},\ \bibinfo {pages} {3457--3466} (\bibinfo {year}
  {2005})}\BibitemShut {NoStop}%
\end{thebibliography}
%

\end{document}